\documentclass[modern]{aastex63} % Used this one for astro-ph

\shorttitle{Stellar Diameters from the NPOI}
\shortauthors{Baines et al.}

\begin{document}

% -------------------------------------------------------------------
% -------------------------------------------------------------------
% -------------------------------------------------------------------
% ----------------- TAKE THIS OUT BEFORE SUBMISSION -----------------
% Distribution Statement A: Approved for public release, distribution is unlimited.
% Add as a footnote to my affiliation to make it appear at the bottom of the page.
% -------------------------------------------------------------------
% -------------------------------------------------------------------
% -------------------------------------------------------------------
% -------------------------------------------------------------------

\title{Angular Diameters and Fundamental Parameters of Forty-Four Stars from the Navy Precision Optical Interferometer}

\author{Ellyn K. Baines}
\affil{Naval Research Laboratory, Remote Sensing Division, 4555 Overlook Ave SW, Washington, DC 20375, USA}
\email{ellyn.baines@nrl.navy.mil}

\author{J. Thomas Armstrong}
\affil{Computational Physics Inc., 8001 Braddock Rd, Suite 210, Springfield, VA 22151, USA}

\author{James H. Clark III}
\affil{Naval Research Laboratory, Remote Sensing Division, 4555 Overlook Ave SW, Washington, DC 20375, USA}

\author{Jim Gorney}
\affil{Lowell Observatory, 1400 W. Mars Hill Rd, Flagstaff, AZ 86001, USA}

\author{Donald J. Hutter}
\affil{Central Michigan University, Department of Physics, Mount Pleasant, MI 48859, USA}

\author{Anders M. Jorgensen}
\affil{New Mexico Institute of Mining and Technology, 801 Leroy Place, Socorro, NM 87801, USA}

\author{Casey Kyte}
\affil{Lowell Observatory, 1400 W. Mars Hill Rd, Flagstaff, AZ 86001, USA} 

\author{David Mozurkewich}
\affil{Seabrook Engineering, 9310 Dubarry Ave, Seabrook, MD 20706, USA}

\author{Ishara Nisley}
\affil{Lowell Observatory, 1400 W. Mars Hill Rd, Flagstaff, AZ 86001, USA} 

\author{Jason Sanborn}
\affil{Lowell Observatory, 1400 W. Mars Hill Rd, Flagstaff, AZ 86001, USA} 

\author{Henrique R. Schmitt}
\affil{Naval Research Laboratory, Remote Sensing Division, 4555 Overlook Ave SW, Washington, DC 20375, USA}

\author{Gerard T. van Belle}
\affil{Lowell Observatory, 1400 W. Mars Hill Rd, Flagstaff, AZ 86001, USA} 

\begin{abstract}

We measured the angular diameters of 44 stars with the Navy Precision Optical Interferometer, obtaining uncertainties on the limb darkened diameter of 2$\%$ or less for all but four stars. We then used our diameters with \emph{Gaia} or \emph{Hipparcos} parallaxes to calculate each star's physical radius. We gathered information from the literature to determine bolometric flux and luminosity, and combined that with our diameters to produce an effective temperature. Our sample consists of mostly giant stars, and spans a wide range of spectral classes from B to M.

\end{abstract}

\keywords{stars: fundamental parameters, techniques: high angular resolution, techniques: interferometric}

%%%%%%%%%%%%%%%%%%%%%%%%%%%%% Introduction %%%%%%%%%%%%%%%%%%%%%%%%%%%%%

\section{Introduction} 

One of interferometry's greatest strengths is its remarkable resolution, which is an order of magnitude better than the world's largest telescopes equipped with adaptive optics \citep{2020MNRAS.493.2377R}. Interferometric angular diameter measurements are the bedrock on which fundamental stellar parameters are based, presenting a direct measurement with no dependence on stellar models. It works across a range of stellar types, from young stars \citep[e.g.,][]{2008SPIE.7013E..0UT} to collections of main-sequence dwarfs \citep[e.g.,][]{2012ApJ...757..112B} to giants and supergiants \citep[e.g.,][]{2016AJ....152...66B, 2017AandA...597A...9W}. Interferometric measurements help to more completely describe interesting subclasses such as O stars \citep{2018ApJ...869...37G}, exoplanet hosts \citep[e.g.,][]{2016ApJ...822L...3J, 2018MNRAS.477.4403W}, asteroseismic targets  \citep[e.g.,][]{2014ApJ...781...90B}, carbon stars \citep{2013ApJ...775...45V}, and so on. 

The ability to obtain high precision stellar diameters is important for many reasons, and one of the most timely applications has to do with characterizing exoplanets. We have seen how the combination of transit measurements from \emph{Kepler} or \emph{TESS} and ground-based radial velocity measurements provide insight into the broad population of exoplanets \citep[e.g.,][]{2014ApJS..210...20M, 2018AJ....155..158B, 2021ApJS..255....6D}, but the results are fundamentally plagued by one major uncertainty: how big are the host stars? The planetary diameter measured through transit observations is relative to the stellar diameter, so the uncertainty in stellar size translates directly to uncertainty in planetary size \citep{2011AJ....142..176M, 2017ApJ...845...65N}. 

Once we have interferometric diameters with precisions of 1-2$\%$, we can better characterize both the star and its planet through model-independent determinations of their masses and densities. This represents a 4- to 6-fold improvement in the density uncertainty for a newly discovered exoplanet, which can have meaningful implications for the type of planet it is. This is essential with the \emph{Kepler} and \emph{TESS} missions finding so many fascinating planetary systems.

Interferometric measurements also have a vital role to play in testing stellar models and acting as benchmarks for large stellar surveys \citep[e.g.,][]{2020AandA...642A.101P, 2020AandA...640A..25K}. This is particularly true for today's high signal-to-noise stellar spectroscopic studies and the \emph{Gaia} survey, where a reliable collection of fundamentally calibrated stellar effective temperatures based on accurate stellar diameters is needed \citep{2020MNRAS.493.2377R}. Additionally, interferometry has been used to demonstrate discrepancies between predicted outcomes from models and what is observed in terms of the model overestimating temperatures and underestimating radii \citep{2012ApJ...757..112B}.

The work described here is a continuation of the survey project in \citet{2018AJ....155...30B}, where we presented the angular diameters and other fundamental stellar properties for 87 stars. This paper is organized as follows: Section 2 discusses the instrument and the observing process, including the selection and characterization of calibrator stars; Section 3 describes the visibility measurements and how we determined various stellar parameters, including the radius, bolometric flux, extinction, luminosity, and effective temperature for each target; and Section 4 provides notes on individual stars, when required, and what our next steps will be.

%\clearpage

%%%%%%%%%%%%%%%%%%%%%%%%% Interferometric observations %%%%%%%%%%%%%%%%%%%%

\section{Interferometric Observations}

\subsection{The Navy Precision Optical Interferometer}

The Navy Precision Optical Interferometer (NPOI) is an optical interferometer located on Anderson Mesa, AZ (see Armstrong et al. 1998 for for the instrument description and Hummel et al. 2003 and Benson, Hummel, \& Mozurkewich 2003 for details about the beam combiner). The NPOI consists of two nested arrays: the four fixed stations of the astrometric array (AC, AE, AW, and AN, which stand for astrometric center, east, west, and north, respectively) that are clustered at the center of the array, and the stations of the imaging array. The latter are arranged along three arms with general north, east, and west orientations. Each arm has ten piers where a siderostat can be installed, which means the imaging array can be reconfigured as needed. 

This paper includes data from the four astrometric stations and seven of the imaging stations: E2, E3, E4, E6, E7, W4, and W7. The stations are labeled according to which arm they are on and how far away they are from the array center, with 1 being closest and 10 being farthest away. This produces baselines from 7.8 m (E2-E4) to 97.5 m (E7-W7). 

The NPOI uses a 12.5-cm diameter region of a 50-cm siderostat in both the astrometric and imaging stations. We can combine light from any of the astrometric and imaging stations that are appropriate for our science goals, up to six stations at a time. The current magnitude limit is $\sim$5.5 in the $V$-band under normal conditions and $\sim$6.0 in excellent conditions.\footnote{The NPOI has received three 1-meter telescopes, which will improve the array's sensitivity to $V$ = 9 when they are fully integrated into the system.}

We observed the 44 targets from 1997 to 2021, a data set that totals over 85,000 calibrated data points. The stars observed in 2020 and 2021 were selected for their large angular sizes ($\geqslant$6 mas in 2020 and $\geqslant$3 mas in 2021) due to the configuration available at the time that consisted of short baseline(s). The other targets include data previously unpublished from the NPOI archive. Table 1 lists each star's identifiers, spectral type, $V$ magnitude, parallax, and metallicity.

Table 2 lists the stars observed, the calibrators, dates, baselines used, and number of data points per night. For the data collected from 2004 to 2021, we used the ``Classic'' beam combiner that takes data over 16 spectral channels spanning 550 to 850 nm \citep{2003AJ....125.2630H, 2016ApJS..227....4H}. Each observation consisted of a 30-second coherent (on the fringe) scan where the fringe contrast was measured every 2 ms. Every coherent scan was paired with an incoherent (off the fringe) scan that was used to estimate the additive bias affecting fringe measurements \citep{2003AJ....125.2630H}. Scans were taken on one to five baselines simultaneously, depending on the array configuration available at the time. Each coherent scan was averaged to 1-second data points, and then again to a single 30-second average. The dispersion of the 1-second data points served as an estimate of the internal uncertainties.

For the data obtained from 1997 to 2001, we used the original version of the``Classic'' beam combiner, which recorded data for up to three baselines individually on three spectrographs, dispersing the light into 32 channels in the wavelength range 450 to 950 nm. The data reduction follows procedures described in \citet{1998AJ....116.2536H}. Unlike the observations completed after 2001, each observation includes only a 90-second coherent scan with no corresponding incoherent scan. In this case we estimate the additive bias based on the fringe power at a modulation frequency higher than the one corresponding to the fringe. 

The NPOI's data reduction package $OYSTER$ was developed by C. A. Hummel\footnote{www.eso.org/$\sim$chummel/oyster/oyster.html} and automatically edits data using the method described in \citet{2003AJ....125.2630H}. In addition to that process, we edited out individual data points and/or scans that showed large scatter, on the order of 5-$\sigma$ or higher. This large scatter was more common in the channels corresponding to the short wavelengths, a long-standing feature in NPOI data due to narrower spectral channels, more pronounced atmospheric effects, and lower quantum efficiency of the avalanche photodiode detectors at these shorter wavelengths. Removing those short-wavelength points did not affect the diameter measurements. 

Starting in 2005, the NPOI regularly measures the central wavelengths of all the spectrometer channels in a Fourier transform spectrometer mode in order to characterize the stability of the wavelength scale calibration \citep{2016ApJS..227....4H}. The measurements show the central wavelengths are stable with a 0.6 nm (0.1$\%$) scatter. For data prior to 2005, we incorporated a $\pm$0.5$\%$ error in the wavelength scale. Only five stars in this study include data from 2004, and of those stars only one (HD 216131, $\mu$ Peg) used only 2004 data and had an uncertainty $<0.5\%$. We therefore assigned a 0.5$\%$ error to its diameter to account for the uncertainty in the wavelength scale.

\subsection{Selection and Characterization of Calibrator Stars}

Atmospheric turbulence and instrumental imperfections reduce the signal strength in interferometric observations. Considering that the fringe contrast due to a small star disk\footnote{Here, ``small'' means that the star's diameter is significantly less than the resolution of the interferometer.} is nearly unity and is only weakly dependent on the diameter, we selected small stars to calibrate the atmospheric and instrumental variations out of the final calculations. Calibrator stars and science targets were observed alternately. The observations taken during a given night were obtained using the same configuration, and the time between stars was generally on the order of a few minutes to 10 minutes.

To estimate the calibrator stars' angular diameters, we created spectral energy distribution (SED) fits based on published $UBVRIJHK$ photometric values. We used plane-parallel model atmospheres \citep{2003IAUS..210P.A20C} based on effective temperature ($T_{\rm eff}$), surface gravity (log~$g$), and $E(B-V)$. Stellar models were fit to observed photometry after converting the magnitudes to fluxes using \citet[][for $UBVRI$]{1996AJ....112..307C} and \citet[][for $JHK$]{2003AJ....126.1090C}. Table 3 lists the photometry, $T_{\rm eff}$, log~$g$, and $E(B-V)$ used, and the resulting angular diameters. This is a relatively simple SED fit, unlike the more sophisticated one described in Section 3.2 that we used for the target stars. For calibrator stars, it is an appropriate method, given the insensitivity of the target's measured angular diameter with respect to the calibrator's diameter \citep{2018AJ....155...30B}.

We checked every calibrator star for binarity, variability, and rapid rotation. Some of the calibrator stars chosen featured one or more of those properties, but not to an extent that would affect the calibration process. Any binary separation was beyond the detection limit of the configuration used, while oblateness due to rapid rotation and/or variability did not introduce a variation in the diameter of the star that would be large enough to cause significant calibration issues.

%%%%%%%%%%%%%%%%%%%%%%%% Angular diameter determinations %%%%%%%%%%%%%%%%%

\section{Results}

\subsection{Angular Diameter Measurement}

Interferometric diameter measurements use visibility squared ($V^2$). For a point source, $V^2$ is 1 and it is considered completely unresolved, while star is completely resolved when its $V^2$ reaches zero. For a uniformly-illuminated disk, $V^2 = [2 J_1(x) / x]^2$, where $J_1$ is the Bessel function of the first order, $x = \pi B \theta_{\rm UD} \lambda^{-1}$, $B$ is the projected baseline toward the star's position, $\theta_{\rm UD}$ is the apparent uniform disk angular diameter of the star, and $\lambda$ is the effective wavelength of the observation \citep{1992ARAandA..30..457S}. $\theta_{\rm UD}$ results for our program stars are listed in Table 4. The data are freely available in OIFITS form \citep{2017AandA...597A...8D} upon request.

A more realistic model of a star's disk includes limb darkening (LD).  If a linear LD coefficient $\mu_\lambda$ is used, then
\begin{eqnarray}
V^2 = \left( {1-\mu_\lambda \over 2} + {\mu_\lambda \over 3} \right)^{-1} 
\times 
\left[(1-\mu_\lambda) {J_1(x_{\rm LD}) \over x_{\rm LD}} + \mu_\lambda {\left( \frac{\pi}{2} \right)^{1/2} \frac{J_{3/2}(x_{\rm LD})}{x_{\rm LD}^{3/2}}} \right]^2 .
\end{eqnarray}
where $x_{\rm LD} = \pi B\theta_{\rm LD}\lambda^{-1}$ \citep{1974MNRAS.167..475H}. We used $T_{\rm eff}$, log $g$, and metallicity ([Fe/H]) values from the literature with an assumed microturbulent velocity of 2 km s$^{\rm -1}$ to obtain $\mu_\lambda$ from \citet{2011AandA...529A..75C}. We used the ATLAS stellar model in the \emph{R}-band, the waveband most closely matched to the central wavelength of the NPOI's bandpass. A more sophisticated analysis of these stars would include how limb darknening depends on wavelength and is non-linear (see Section 4). Still, the treatment here is a fair approach, given that the strength of the limb darkening for star is related to the height of the second maximum of the visibility curve \citep{{2001AandA...377..981W}} and we deal largely with measurements before the first minimum.

The $T_{\rm eff}$, log $g$, [Fe/H], and $\mu_\lambda$ used and the resulting limb darkened diameters ($\theta_{\rm LD}$) are listed in Table 4 along with the maximum spatial frequency for each star's data set, and the number of data points in the angular diameter fit. Figure 1 shows the $\theta_{\rm LD}$ fit for HD 3627 ($\delta$ And) as an example. The remaining plots are included in the supplementary material of the \emph{Astronomical Journal}. 

% COME BACK TO PREVIOUS PARAGRAPH ABOUT TABLE

We used the procedure described in \citet{2018AJ....155...30B} to estimate the uncertainty for the angular diameter, which we summarize here: if we use only the collected data points without regard to correlations within a scan, the diameter's uncertainty can be significantly underestimated. Instead we used a modified bootstrap Monte Carlo method developed by \citet{2010SPIE.7734E.103T} to create a large number of synthetic datasets by selecting entire scans at random. The width of the distribution of diameters fit to these datasets becomes our measure of the uncertainty for the diameter (see Figure \ref{plot_gauss}). 

\subsection{Stellar Radius, Luminosity, and Effective Temperature}

When available, the parallax from the \citet{2020yCat.1350....0G} was converted into a distance and combined with our measured diameters to calculate the physical radius $R$. Otherwise, parallaxes from \citet{2007AandA...474..653V}, \citet{2016AandA...593A..51M}, and \citet{2018AandA...616A...1G} were used, which was the case for eight stars. 

In order to determine each star's luminosity ($L$) and $T_{\rm eff}$, we created SED fits using photometric values published in \citet{1966CoLPL...4...99J}, \citet{1975RMxAA...1..299J}, \citet{1984AandAS...57..357O}, \citet{1988iras....7.....H}, \citet{1988iras....1.....B}, \citet{1993cio..book.....G}, \citet{1993AandAS..100..591O}, \citet{1999yCat.2225....0G}, \citet{2000AandA...355L..27H}, \citet{2002yCat.2237....0D}, \citet{2002AandA...384..180F}, \citet{2003yCat.2246....0C}, \citet{2004ApJS..154..673S}, \citet{2006MNRAS.369..723N}, \citet{2009ApJ...694.1085V}, \citet{2012MNRAS.419.1637L}, and \citet{2012yCat.1322....0Z}. The assigned uncertainties for the 2MASS infrared measurements are as reported in \citet{2003yCat.2246....0C}, and an uncertainty of 0.05 mag was assigned to the optical measurements.  

We determined the best fit stellar spectral template to the photometry from the flux-calibrated stellar spectral atlas of \citet{1998PASP..110..863P} using the $\chi^2$ minimization technique \citep{1992nrca.book.....P, 2003psa..book.....W}. This produced the bolometric flux ($F_{\rm BOL}$) for each star and allowed for the calculation of extinction ($A_{\rm V}$) with the wavelength-dependent reddening relations of \citet{1989ApJ...345..245C}.

We combined our $F_{\rm BOL}$ values with the stars' distances to estimate $L$ using $L = 4 \pi d^2 F_{\rm BOL}$. We also combined the $F_{\rm BOL}$ with $\theta_{\rm LD}$ to determine each star's effective temperature by inverting the relation,
\begin{equation}
F_{\rm BOL} = {1 \over 4} \theta_{\rm LD}^2 \sigma T_{\rm eff}^4,
\end{equation}
where $\sigma$ is the Stefan-Boltzmann constant and $\theta_{\rm LD}$ is in radians \citep{1999AJ....117..521V}. The resulting $R$, $F_{\rm BOL}$, $A_{\rm V}$, $T_{\rm eff}$, and $L$ are listed in Table 5.

Because $\mu_\lambda$ is chosen based on a given $T_{\rm eff}$, we used an iterative process to determine the final $\theta_{\rm LD}$. We started with the initial $\theta_{\rm LD}$, calculated $T_{\rm eff}$, and used that new $T_{\rm eff}$ to see if $\mu_\lambda$ changed. For 16 stars, the $\theta_{\rm LD}$ did not change during this process. Generally, $\mu_\lambda$ changed by an average of 0.02, and the largest difference was 0.11. The resulting $\theta_{\rm LD}$ values changed at most by 1.7$\%$, and the average difference was 0.3$\%$ (0.013 mas). 

Overall, the new $T_{\rm eff}$ differed from the original $T_{\rm eff}$ by an average of 41 K. The largest outlier was HD 35497 ($\beta$ Tau) with a $\Delta T_{\rm eff}$ of 275 K. Even so, this star's final $\theta_{\rm LD}$ and $T_{\rm eff}$ settled after only one iteration like all the other stars in the sample. Table 4 lists the initial $\mu_\lambda$ and the resulting $\theta_{\rm LD}$, as well as the final $\mu_\lambda$ and $\theta_{\rm LD}$. The final $\theta_{\rm LD}$ is little affected by the choice of $\mu_{\lambda}$: a 10$\%$ change in $\mu_{\lambda}$ results in at most a 0.3$\%$ change in $\theta_{\rm LD}$, with an average of a 0.1$\%$ change.

\section{Discussion and Conclusions}

The measurements presented here make two assumptions about the targets: they are single, and they are not rapid rotators. There were only two rapid rotators in the sample with $v$$\sin$$i$ higher than 100 km s$^{\rm -1}$: HD 109387 ($\kappa$ Dra) and HD 129989 ($\epsilon$ Boo A). HD 109387 is the smallest star we measure here at 0.9 mas, so any asymmetries in its shape would be too small for the NPOI to detect. HD 129989 is larger at over 4 mas, though the sampling of the $u-v$ plane for this star does not give us enough coverage to detect asymmetries.

As for binarity, \citet{2016ApJS..227....4H} demonstrated that the NPOI's detection sensitivity spans separations from 3 to 860 mas with magnitude differences of 3.0 for most binary systems, and up to 3.5 when the component spectral types differ by less than two. Taking these limits into account, three stars in our sample have known stellar companions that require further investigation:

\emph{HD 35497 ($\beta$ Tau)}: This star includes a companion at 0''.1 mas with a $\Delta m_V$ of 0 in the Washington Double Star (WDS) catalog \citep{2001AJ....122.3466M}, though the companion was never confirmed. \citet{2006AandA...447..685A} list this target as a single lined spectroscopic binary, which implies a $\Delta m_V$ outside of the NPOI's detection limits. We fit this target as a single star, following the example of \citet{2019ApJ...873...91G}. %% HS: I'm not sure about this one. The increased rms suggests that you may be detecting the binary.

\emph{HD 74442 ($\delta$ Cnc)}: The WDS catalog lists this system as Aa-Ab with a separation of 0''.1 and a $\Delta m_V$ of 0.96, as well as an AB companion with a larger separation and $\Delta m_V$ well out of the NPOI's range. For the inner binary, \citet{2007PASP..119..886E} found a standard deviation in the star's velocity of 0.15 km s$^{-1}$ in 124 observations. This small scatter makes it unlikely that this star is a binary. We follow \citet{2017AJ....153...16S} in treating this star as single.
%\citet{2019MNRAS.490.3158C} cite a separation of 50 mas

%\emph{HD 208816 (HR 8383)}: The WDS lists this as a $\zeta$ Aurigae-type binary with a separation of 0''.1 and no $\Delta m_V$ indicated, and noted that ``visual duplicity uncertain''. \citet{2020RNAAS...4...12P} provides spectral types of M2Ia-Iab+B8V, which would imply a $\Delta m_V$ outside the NPOI's detection limits. We therefore treated this star as single.

\emph{HD 208816 (HR 8383)}: The WDS lists this as a $\zeta$ Aurigae-type binary with a separation of 0''.1 and no $\Delta m_V$ indicated, and noted that ``visual duplicity uncertain''. However, \citet{2020RNAAS...4...12P} provided spectral types of M2Ia-Iab+B8V, and \citet{1971MNRAS.155..203H} showed the two components have similiar masses (18.3 $M_\odot$ and 19.8 $M_\odot$). This implies the binary would affect our measurements, but an earlier study by \citet{1970VA.....12..147W} listed a $\Delta m_V$ of $\sim$3.5, which would be out of the range of the NPOI to detect. We include our measurement here for completeness, though we note the binary could affect the diameter.

The NPOI has searched for binary companions in five additional targets as part of an ongoing multiplicity survey: no close companions were found for HD 9826 = $\upsilon$ And and HD 61421 = $\alpha$ CMi, and no companions were found for HD 19373 = $\iota$ Per \citep{2016ApJS..227....4H, 2019ApJS..243...32H}, HD 109387 = $\kappa$ Dra, or HD 192909 = 32 Cyg \citep{2016ApJS..227....4H}.
%HD 109387 = $\kappa$ Dra (Hutter, private communication)

Four of the 44 stars presented here have not yet been previously measured interferometrically (see Table 6). Figure \ref{lit_diam_compare} shows the comparison between diameters from the literature and those measured here. The diameters generally fall near the 1:1 line, with the largest outlier being HD 198026 (k Aqr). We measured an angular diameter of 7.124$\pm$0.088 mas, while \citet{1996AJ....111.1705D, 1998AJ....116..981D} put it at 5.5$\pm$0.7 mas. Some of the variations between previous measurements and those presented here may be dependent on what limb darkening law, if any, was used to determine the final angular diameter.

%\emph{HD 42995, $\eta$ Gem}: This star has been measured three times previously, with diameters of 11.75$\pm$0.27 mas \citep{1993ApJ...406..215Q} and 11.789$\pm$0.118 mas \citep{2003AJ....126.2502M}, both from the Mark III interferometer, and 12.57$\pm$0.04 mas \citep{2005AandA...434.1201R} from the Very Large Telescope Interferometer (VLTI). Of the three measurements, only 2003 Mark III measurement included limb darkening in the diameter presented. The VLTI measurement used the MIDI instrument in the $N$-band at 8-13 $\mu$m. In Figure \ref{lit_diam_compare}, we compared our measurement of 11.112$\pm$0.024 mas with the most recent of the literature diameters, although the other two are closer to our result and the Mark III's wavelength coverage (451 nm to 800 nm) more nearly matches the NPOI's.

Our next step is to directly measure the limb darkening for the stars with data through or beyond the first null, where $V^2$ drops to zero (e.g., see the visibility plots for HD 60522 = $\upsilon$ Gem, HD 61421 = $\alpha$ CMi, HD 129989 = $\epsilon$ Boo A). Before the first null, the visibility curve is dominated by the star's angular diameter, which is what we measured here. After the first null, second order effects such as limb darkening have more influence, and specific limb darkening models and prescriptions can be directly tested \citep{2001AandA...377..981W}. 

These stars are also good test cases for using the dependence of the spatial frequency ($u_0 = B / \lambda_0$) of the fringe visibility null on wavelength and projected baseline as a probe of stellar atmospheres, as described by \citet{2019JAI.....850012A}. As the Earth rotates, changing the baseline, the wavelength at the null ($\lambda_0$) also changes. For a gray atmosphere, $u_0$ remains constant as the Earth rotates, so $\lambda_0$ varies directly as $B$. Any departure from that variation suggests a variation in limb darkening with wavelength and/or a wavelength-dependent variation in angular size.

\begin{acknowledgements}

This material is based upon work supported by the National Aeronautics and Space Administration under Grant 18-XRP18$\_$2-0017 issued through the Exoplanets Research Program. The Navy Precision Optical Interferometer is a joint project of the Naval Research Laboratory and the U.S. Naval Observatory, in cooperation with Lowell Observatory, and is funded by the Office of Naval Research and the Oceanographer of the Navy. This research has made use of the SIMBAD database and Vizier catalogue access tool, operated at CDS, Strasbourg, France. This publication made use of data products from the Two Micron All Sky Survey, which is a joint project of the University of Massachusetts and the Infrared Processing and Analysis Center/California Institute of Technology, funded by the National Aeronautics and Space Administration and the National Science Foundation. This research has made use of the Jean-Marie Mariotti Center JSDC catalogue, available at http://www.jmmc.fr/catalogue$\_$jsdc.htm.

\end{acknowledgements}

\clearpage

%%%%%%%%%%%%%%%%%%%%%%%%%%%%%% General properties %%%%%%%%%%%%%%%%%%%%%%%%%%%%%%%%%%%%%%%

%\startlongtable 
\begin{deluxetable}{cccccccrcc}
%\tablewidth{0pc}
\tabletypesize{\scriptsize}
\tablecaption{Sample Star Properties.\label{general_properties}}
\tablehead{
 \colhead{}   & \colhead{}   & \colhead{}    & \colhead{Other} & \colhead{Spectral} & \colhead{$V$}   & \colhead{Parallax} & \colhead{}    & \colhead{}       & \colhead{}    \\
 \colhead{HD} & \colhead{HR} & \colhead{FK5} & \colhead{Name}  & \colhead{Type}     & \colhead{(mag)} & \colhead{(mas)}    & \colhead{Ref} & \colhead{[Fe/H]} & \colhead{Ref} \\ }
\startdata
3627   & 165  & 20   & $\delta$ And      & K3 III     & 3.27$\pm$0.03 & 31.57$\pm$0.45  & 1 & 0.15  & 5	\\
4656   & 224  & 28   & $\delta$ Psc      & K5 III     & 4.43$\pm$0.01 & 10.86$\pm$0.18  & 1 & -0.11 & 5	\\
5112   & 248  & 1022 & 20 Cet            & K5 III     & 4.76$\pm$0.01 & 6.08$\pm$0.19   & 1 & 0.00  & 6	\\
6805   & 334  & 40   & $\eta$ Cet        & K2 III     & 3.44$\pm$0.01 & 27.06$\pm$0.18  & 1 & 0.09  & 5	\\
9826   & 458  & 1045 & $\upsilon$ And    & F9 V       & 4.09$\pm$0.01 & 74.19$\pm$0.21  & 1 & 0.08  & 5	\\
10380  & 489  & 56   & $\nu$ Psc         & K3 III     & 4.44$\pm$0.01 & 8.93$\pm$0.16   & 1 & -0.19 & 5	\\
19373  & 937  & 112  & $\iota$ Per       & G0 V       & 4.05$\pm$0.01 & 94.87$\pm$0.23  & 2 & 0.08  & 7	\\
20902  & 1017 & 120  & $\alpha$ Per      & F5 I       & 1.80$\pm$0.01 & 6.44$\pm$0.17   & 2 & 0.14  & 5	\\
25025  & 1231 & 149  & $\gamma$ Eri      & M0 III     & 2.95$\pm$0.01 & 17.00$\pm$0.23  & 1 & 0.00  & 8	\\
28307  & 1411 & -    & $\theta^{1}$ Tau  & G9 III     & 3.84$\pm$0.01 & 24.77$\pm$0.44  & 1 & 0.13  & 5	\\
35497  & 1791 & 202  & $\beta$ Tau       & B7 III     & 1.65$\pm$0.01 & 24.36$\pm$0.34  & 2 & 0.20  & 5	\\
39003  & 2012 & 221  & $\nu$ Aur         & K0.5 III   & 3.96$\pm$0.01 & 16.14$\pm$0.44  & 1 & -0.03 & 5	\\
42995  & 2216 & -    & $\eta$ Gem        & M2 III     & 3.28$\pm$0.01 & 4.73$\pm$1.02   & 3 & 0.00  & 5	\\
60522  & 2905 & 1196 & $\upsilon$ Gem    & M0 III     & 4.06$\pm$0.01 & 12.88$\pm$0.23  & 1 & -0.27 & 5	\\
61421  & 2943 & 291  & $\alpha$ CMi      & F5 IV-V    & 0.37$\pm$0.01 & 284.56$\pm$1.26 & 4 & -0.04 & 9	\\
61935  & 2970 & 293  & $\alpha$ Mon      & G9.5 III   & 3.93$\pm$0.01 & 22.38$\pm$0.13  & 1 & -0.04 & 5	\\
74442  & 3461 & 326  & $\delta$ Cnc      & K0 III     & 3.94$\pm$0.01 & 23.83$\pm$0.19  & 1 & -0.03 & 5	\\
80493  & 3705 & 352  & $\alpha$ Lyn      & K6 III     & 3.14$\pm$0.01 & 14.70$\pm$0.18  & 1 & -0.13 & 5	\\
89758  & 4069 & 386  & $\mu$ UMa         & M0 III     & 3.05$\pm$0.01 & 17.80$\pm$0.39  & 1 & -0.04 & 5	\\
102224 & 4518 & 441  & $\chi$ UMa        & K0.5 III   & 3.71$\pm$0.01 & 16.44$\pm$0.11  & 1 & -0.36 & 5	\\
109387 & 4787 & 472  & $\kappa$ Dra      & B6 III     & 3.85$\pm$0.03 & 7.01$\pm$0.30   & 1 & -0.65 & 5	\\
112300 & 4910 & 484  & $\delta$ Vir      & M3 III     & 3.38$\pm$0.01 & 17.41$\pm$0.25  & 1 & -0.06 & 5	\\
129989 & 5506 & -    & $\epsilon$ Boo A  & K0 II-III  & 2.38$\pm$0.01 & 13.83$\pm$0.49  & 1 & 0.05  & 8	\\
131873 & 5563 & 550  & $\beta$ UMi       & K4 III     & 2.08$\pm$0.01 & 24.91$\pm$0.12  & 2 & -0.17 & 5	\\
132813 & 5589 & 554  & -                 & M4.5 III   & 4.59$\pm$0.02 & 9.85$\pm$0.49   & 1 & 0.00  & 8	\\
133165 & 5601 & 3190 & 110 Vir           & K0 III     & 4.40$\pm$0.01 & 16.75$\pm$0.13  & 1 & -0.21 & 5	\\
135722 & 5681 & 563  & $\delta$ Boo      & G8 III     & 3.48$\pm$0.01 & 27.07$\pm$0.13  & 1 & -0.36 & 5	\\
141477 & 5879 & 584  & $\kappa$ Ser      & M0.5 III   & 4.10$\pm$0.01 & 8.52$\pm$0.17   & 1 & 0.00  & 8	\\
163993 & 6703 & 674  & $\xi$ Her         & G8.5 III   & 3.70$\pm$0.01 & 23.85$\pm$0.11  & 1 & 0.03  & 5	\\
164058 & 6705 & 676  & $\gamma$ Dra      & K5 III     & 2.23$\pm$0.01 & 21.14$\pm$0.10  & 2 & -0.13 & 5	\\
175588 & 7139 & -    & $\delta^{02}$ Lyr & M4 II      & 4.28$\pm$0.02 & 4.24$\pm$0.30   & 1 & -0.04 & 5	\\
189319 & 7635 & 752  & $\gamma$ Sge      & M0 III     & 3.48$\pm$0.03 & 11.34$\pm$0.17  & 1 & -0.08 & 5	\\
192909 & 7751 & -    & 32 Cyg            & K7I-II+B1V & 3.98$\pm$0.01 & 3.26$\pm$0.19   & 1 & -0.33 & 5	\\
197989 & 7949 & 780  & $\epsilon$ Cyg    & K0 III     & 2.46$\pm$0.01 & 43.18$\pm$0.94  & 3 & -0.10 & 5	\\
198026 & 7951 & 1543 & k Aqr             & M3 III     & 4.43$\pm$0.01 & 6.47$\pm$0.21   & 1 & 0.00  & 6	\\
200905 & 8079 & 792  & $\xi$ Cyg         & K4.5 I-II  & 3.73$\pm$0.04 & 2.84$\pm$0.13   & 1 & -0.26 & 5	\\
208816 & 8383 & 3756 & -                 & M2I+B8V    & 4.94$\pm$0.06 & 1.00$\pm$0.11   & 1 & 0.01  & 5	\\
210745 & 8465 & 836  & $\zeta$ Cep       & K1.5 I     & 3.35$\pm$0.01 & 3.30$\pm$0.15   & 1 & 0.04  & 5	\\
213306 & 8571 & 847  & $\delta$ Cep      & F5I+B7-8   & 3.56$\pm$0.10 & 3.56$\pm$0.15   & 1 & 0.09  & 10	\\
214868 & 8632 & -    & 11 Lac            & K2.5 III   & 4.48$\pm$0.02 & 9.32$\pm$0.11   & 1 & -0.14 & 5	\\
216131 & 8684 & 862  & $\mu$ Peg         & G8III      & 3.49$\pm$0.01 & 28.93$\pm$0.19  & 1 & -0.05 & 5	\\
216386 & 8698 & 864  & $\lambda$ Aqr     & M2.5 III   & 3.75$\pm$0.03 & 8.94$\pm$0.24   & 1 & 0.00  & 6	\\
218452 & 8804 & 3852 & 4 And             & K5 III     & 5.32$\pm$0.01 & 9.16$\pm$0.08   & 1 & 0.07  & 5	\\
224935 & 9089 & 1630 & 30 Psc            & M3 III     & 4.40$\pm$0.00 & 7.88$\pm$0.41   & 1 & 0.00  & 6	\\
\enddata
\tablecomments{Spectral types are from SIMBAD, $V$ magnitudes are from \citet{Mermilliod}, parallaxes and [Fe/H] are from the following sources: 1. \citet{2020yCat.1350....0G}; 2. \citet{2007AandA...474..653V}; 3. \citet{2018AandA...616A...1G}; 4. \citet{2016AandA...593A..51M}; 5. \citet{2012AstL...38..331A}; 6. no [Fe/H] was available so 0.00 was used; 7. \citet{2017ApJ...836...77Y}; 8. \citet{2020AandA...633A..34C}; 9. \citet{2018AandA...619A.134H}; and 10. \citet{2016MNRAS.459.1170S}.}
\end{deluxetable}

\clearpage

%%%%%%%%%%%%%%%%%%%%%%%%%%%%%% Observing Log & Calib Info %%%%%%%%%%%%%%%%%%%%%%%%%%%%%%%%%%%%%%%

\begin{deluxetable}{ccllc}
\tablewidth{0pc}
%\tabletypesize{\scriptsize}
\tablecaption{Observing Log.\label{observations}}
\tablehead{
 \colhead{Target}  & \colhead{Calibrator} & \colhead{Date} & \colhead{Baselines}      & \colhead{$\#$} \\
 \colhead{HD}      & \colhead{HD}         & \colhead{(UT)} & \colhead{Used$^\dagger$} & \colhead{Data Points} \\ }
\startdata
3627   & 5448   & 1998 Nov 19 & AW-E4                             & 114 \\
       &        & 1999 Jan 08 & AW-E2, AW-E4, E2-E4               & 51  \\
4656   & 886    & 2012 Nov 27 & AC-AE, AC-AW, AC-E6, AW-E6        & 259 \\      
       &        & 2012 Nov 29 & AC-AE, AC-AW, AC-E6, AW-E6        & 222 \\
       &        & 2012 Nov 30 & AC-AE                             & 50  \\     
5112   & 7804   & 2012 Oct 27 & AC-AE, AC-W7, AE-W7               & 156 \\
       &        & 2012 Oct 28 & AC-AE, AC-W7, AE-W7               & 90  \\
       &        & 2012 Oct 29 & AC-AE, AC-W7, AE-W7               & 117 \\
       &        & 2012 Nov 04 & AC-AE, AC-E6, AC-W7, AE-W7, E6-W7 & 114 \\
6805   & 11171  & 2012 Nov 07 & AC-AE, AC-E6, AC-W7, AE-W7, E6-W7 & 385 \\
       &        & 2012 Nov 12 & AC-AE, AC-E6, AC-W7, AE-W7, E6-W7 & 185 \\
       &        & 2012 Nov 13 & AC-AE, AC-W7, AE-W7               & 245 \\
       &        & 2012 Nov 14 & AC-AE, AC-E6, AC-W7, AE-W7, E6-W7 & 522 \\
       &        & 2012 Nov 15 & AC-AE, AC-E6, AC-W7, AE-W7, E6-W7 & 400 \\
       &        & 2012 Nov 17 & AC-AE, AC-E6, AC-W7, AE-W7, E6-W7 & 385 \\
       &        & 2012 Nov 21 & AC-AE, AC-E6, AC-W7, AE-W7, E6-W7 & 288 \\   
       &        & 2012 Nov 26 & AC-AE, AC-E6, AC-W7, AE-W7, E6-W7 & 576 \\
9826   & 5448   & 2017 Nov 10 & AC-AE, AC-AW, AC-E6, AW-E6        & 1020 \\
       &        & 2017 Nov 16 & AC-AE, AC-AW, AC-E6, AW-E6        & 480 \\
       &        & 2017 Nov 22 & AC-AE, AC-AW, AC-E6, AW-E6        & 781 \\
       &        & 2017 Nov 23 & AC-AE, AC-AW, AC-E6, AW-E6        & 495 \\
       &        & 2017 Dec 03 & AC-AE, AC-AW, AC-E6, AW-E6        & 647 \\
       &        & 2017 Dec 13 & AC-AE, AC-AW, AC-E6, AW-E6        & 245 \\
       &        & 2017 Dec 14 & AC-AE, AC-AW, AC-E6, AW-E6        & 354 \\
       &        & 2017 Dec 28 & AC-AE, AC-AW, AC-E6, AW-E6        & 187 \\
\enddata
\tablecomments{$^\dagger$The maximum baseline lengths are AC-AE 18.9 m, AC-AW 22.2 m, AC-E6 34.4 m, AC-W7 51.3 m, AE-W7 64.2 m, AW-E2 31.9 m, AW-E4 14.0 m, AW-E6 53.3 m, E2-E4 7.8 m, and E6-W7 79.4 m. This table shows the information for several stars as an example; the full table is available on the electronic version of the \emph{Astronomical Journal}.}
\end{deluxetable}

\clearpage

%%%%%%%%%%%%%%%%%%%%%%%%%%%%% Calibrators %%%%%%%%%%%%%%%%%%%%%%%%%%%%%%%%%%%

%\startlongtable 
%\begin{longrotatetable}
\begin{deluxetable}{lccccccccccccccc}
%\rotate
%\tablewidth{0pc}
\tabletypesize{\scriptsize}
\tablecaption{Calibrator Stars' SED Inputs and Angular Diameters. \label{calibrators}}
\tablehead{
 \colhead{ }  & \colhead{Spec} & \colhead{$U$}   & \colhead{$B$}   & \colhead{$V$}   & \colhead{$R$}   & \colhead{$I$}   & \colhead{$J$}   & \colhead{$H$}   & \colhead{$K$}   & \colhead{$T_{\rm eff}$} & \colhead{log $g$}     & \colhead{}    & \colhead{}         & \colhead{}    & \colhead{$\theta_{\rm est}$} \\
 \colhead{HD} & \colhead{Type} & \colhead{(mag)} & \colhead{(mag)} & \colhead{(mag)} & \colhead{(mag)} & \colhead{(mag)} & \colhead{(mag)} & \colhead{(mag)} & \colhead{(mag)} & \colhead{(K)}           & \colhead{(cm s$^{-2}$)} & \colhead{Ref} & \colhead{$E(B-V)$} & \colhead{Ref} & \colhead{(mas)}   }
\startdata
886    & B2 IV       & 1.75 & 2.61 & 2.83 & 2.88 & 3.06 & 3.50 & 3.64 & 3.77 & 21944 & 3.93 & 1 & 0.02 & 5  & 0.45$\pm$0.02	\\
5448   & A6 V        & 4.14 & 3.99 & 3.86 & 3.78 & 3.71 & 3.62 & 3.65 & 3.64 & 8128  & 3.78 & 2 & 0.00 & 6  & 0.69$\pm$0.03	\\
7804   & A1 V        & 5.34 & 5.23 & 5.16 & 5.10 & 5.08 & 5.19 & 5.04 & 4.92 & 8710  & 4.00 & 2 & 0.02 & 6  & 0.35$\pm$0.02	\\
11171  & F0 V        & 5.04 & 5.00 & 4.67 & 4.47 & 4.31 & 3.66 & 3.47 & 3.87 & 7244  & 4.24 & 2 & 0.02 & 6  & 0.66$\pm$0.03	\\
24760  & B1.5 III    & 1.74 & 2.69 & 2.89 & 2.95 & 3.11 & 3.45 & 3.60 & 3.71 & 26405 & 3.85 & 3 & 0.09 & 7  & 0.35$\pm$0.02	\\
26574  & F0 III      & 4.53 & 4.37 & 4.04 & 3.84 & 3.68 & 3.43 & 3.25 & 3.21 & 7079  & 3.66 & 2 & 0.02 & 6  & 0.86$\pm$0.04	\\
29248  & B2 III      & 2.84 & 3.72 & 3.93 & 4.02 & 4.20 & 4.08 & 4.58 & 4.51 & 20000 & 3.4  & 4 & 0.03 & 8  & 0.31$\pm$0.02	\\
29388  & A6 V        & 4.52 & 4.39 & 4.27 & 4.19 & 4.13 & 4.12 & 4.08 & 4.11 & 8128  & 3.88 & 2 & 0.03 & 8  & 0.57$\pm$0.03	\\
32630  & B3 V        & 2.33 & 2.99 & 3.17 & 3.21 & 3.35 & 3.61 & 3.76 & 3.86 & 14125 & 3.94 & 2 & 0.02 & 7  & 0.52$\pm$0.03	\\
35468  & B2 V        & 0.55 & 1.41 & 1.64 & 1.71 & 1.84 & 2.15 & 2.36 & 2.38 & 21380 & 3.81 & 2 & 0.02 & 8  & 0.82$\pm$0.04	\\
40312  & A0 V        & 2.40 & 2.56 & 2.65 & 2.64 & 2.70 & 2.38 & 2.39 & 2.40 & 10715 & 3.60 & 2 & 0.00 & 7  & 0.94$\pm$0.05	\\
50019  & A2 IV       & 3.84 & 3.70 & 3.60 & 3.53 & 3.48 & 3.25 & 3.23 & 3.16 & 8128  & 3.50 & 2 & 0.03 & 6  & 0.83$\pm$0.04	\\
56537  & A4 IV       & 3.79 & 3.69 & 3.58 & 3.50 & 3.44 & 3.54 & 3.50 & 3.54 & 8511  & 4.10 & 2 & 0.02 & 6  & 0.75$\pm$0.04	\\
58715  & B8 V        & 2.51 & 2.80 & 2.89 & 2.87 & 2.94 & 3.06 & 3.11 & 3.10 & 11220 & 3.73 & 2 & 0.02 & 9  & 0.76$\pm$0.04	\\
58946  & F1 V        & 4.47 & 4.49 & 4.18 & 4.00 & 3.84 & 3.22 & 3.16 & 2.98 & 7244  & 4.26 & 2 & 0.00 & 10 & 0.83$\pm$0.04	\\
71155  & A0 V        & 3.86 & 3.88 & 3.90 & 3.89 & 3.92 & 4.12 & 4.09 & 4.08 & 9772  & 4.16 & 2 & 0.03 & 7  & 0.54$\pm$0.03	\\
76756  & A7 V        & 4.58 & 4.40 & 4.26 & 4.18 & 4.11 & 3.98 & 4.03 & 3.94 & 7943  & 3.73 & 2 & 0.06 & 11 & 0.60$\pm$0.03	\\
87696  & A7 V        & 4.74 & 4.67 & 4.49 & 4.38 & 4.29 & 4.27 & 4.05 & 4.00 & 7943  & 4.27 & 2 & 0.01 & 6  & 0.56$\pm$0.03	\\
89021  & A1 IV       & 3.54 & 3.48 & 3.45 & 3.40 & 3.40 & 3.44 & 3.46 & 3.42 & 8913  & 3.84 & 2 & 0.01 & 7  & 0.74$\pm$0.04	\\
95418  & A1 IV       & 2.35 & 2.35 & 2.37 & 2.34 & 2.34 & 2.27 & 2.36 & 2.29 & 8913  & 3.82 & 2 & 0.00 & 7  & 1.23$\pm$0.06	\\
106591 & A2 V        & 3.46 & 3.39 & 3.31 & 3.24 & 3.21 & 3.32 & 3.31 & 3.10 & 8710  & 4.12 & 2 & 0.00 & 6  & 0.81$\pm$0.04	\\
109387 & B6 III      & 3.15 & 3.73 & 3.85 & 3.92 & 4.03 & 3.82 & 3.91 & 3.82 & 14380 & 3.15 & 1 & 0.02 & 9  & 0.44$\pm$0.02	\\
112413 & A0 V        & 2.45 & 2.78 & 2.89 & 2.88 & 2.94 & 3.06 & 3.13 & 3.15 & 12589 & 4.23 & 2 & 0.01 & 5  & 0.67$\pm$0.03	\\
116842 & A5V+M3-4V   & 4.26 & 4.18 & 4.01 & 3.90 & 3.81 & 3.29 & 3.30 & 3.15 & 8128  & 4.18 & 2 & 0.02 & 6  & 0.74$\pm$0.04	\\
118098 & A2 V        & 3.60 & 3.49 & 3.37 & 3.31 & 3.25 & 3.26 & 3.15 & 3.22 & 8511  & 4.19 & 2 & 0.02 & 11 & 0.81$\pm$0.04	\\
122408 & A2 IV/V     & 4.48 & 4.36 & 4.25 & 4.19 & 4.14 & 4.21 & 4.11 & 4.09 & 8128  & 3.58 & 2 & 0.11 & 12 & 0.64$\pm$0.03	\\
125162 & A0 V        & 4.32 & 4.26 & 4.18 & 4.13 & 4.10 & 3.98 & 4.03 & 3.91 & 8710  & 4.26 & 2 & 0.01 & 13 & 0.56$\pm$0.03	\\
128167 & F4 V        & 4.75 & 4.83 & 4.47 & 4.27 & 4.10 & 3.56 & 3.46 & 3.34 & 6918  & 4.37 & 2 & 0.00 & 10 & 0.76$\pm$0.04	\\
130109 & A0 III      & 3.71 & 3.73 & 3.74 & 3.72 & 3.74 & 3.68 & 3.63 & 3.65 & 9550  & 4.07 & 2 & 0.01 & 6  & 0.59$\pm$0.03	\\
141003 & A2 IV       & 3.82 & 3.73 & 3.67 & 3.60 & 3.57 & 3.44 & 3.54 & 3.55 & 8511  & 3.69 & 2 & 0.02 & 7  & 0.73$\pm$0.04	\\
143894 & A3 V        & 4.97 & 4.89 & 4.83 & 4.78 & 4.76 & 5.01 & 4.66 & 4.62 & 8913  & 4.11 & 2 & 0.02 & 6  & 0.40$\pm$0.02	\\
147394 & B5 IV       & 3.19 & 3.74 & 3.90 & 3.94 & 4.07 & 3.93 & 4.09 & 4.29 & 14791 & 3.98 & 2 & 0.03 & 7  & 0.38$\pm$0.02	\\
166014 & B9.5 III    & 3.77 & 3.81 & 3.84 & 3.85 & 3.89 & 3.97 & 3.96 & 3.95 & 10590 & 4.17 & 3 & 0.02 & 9  & 0.52$\pm$0.03	\\
176437 & B9 III      & 3.10 & 3.20 & 3.25 & 3.24 & 3.28 & 3.12 & 3.23 & 3.12 & 10500 & 3.4  & 4 & 0.02 & 5  & 0.71$\pm$0.04	\\
177724 & A0 IV-V     & 3.01 & 3.00 & 2.99 & 2.94 & 2.94 & 3.08 & 3.05 & 2.88 & 9333  & 4.09 & 2 & 0.05 & 7  & 0.91$\pm$0.05	\\
184006 & A5 V        & 4.07 & 3.93 & 3.78 & 3.69 & 3.62 & 3.74 & 3.69 & 3.60 & 7943  & 3.77 & 2 & 0.00 & 6  & 0.70$\pm$0.04	\\
192696 & A3 IV-V     & 4.49 & 4.41 & 4.30 & 4.22 & 4.17 & 4.28 & 4.17 & 4.08 & 8318  & 3.89 & 2 & 0.04 & 6  & 0.57$\pm$0.03	\\
195810 & B6 IV       & 3.44 & 3.91 & 4.03 & 4.07 & 4.18 & 4.66 & 4.55 & 4.38 & 14355 & 3.71 & 3 & 0.02 & 9  & 0.39$\pm$0.02	\\
198001 & B9.5 V      & 3.81 & 3.77 & 3.77 & 3.76 & 3.77 & 3.85 & 3.67 & 3.74 & 9120  & 3.55 & 2 & 0.02 & 5  & 0.63$\pm$0.03	\\
200761 & A1 V        & 4.06 & 4.05 & 4.06 & 4.07 & 4.09 & 4.37 & 4.32 & 4.10 & 9550  & 4.01 & 2 & 0.01 & 6  & 0.50$\pm$0.03	\\
202850 & A0 I        & 3.97 & 4.36 & 4.23 & 4.20 & 4.17 & 3.97 & 3.86 & 3.68 & 11839 & 2.05 & 3 & 0.20 & 7  & 0.55$\pm$0.03	\\
202904 & B2 V        & 3.51 & 4.29 & 4.39 & 4.42 & 4.53 & 4.70 & 4.54 & 4.48 & 20900 & 3.9  & 4 & 0.13 & 14 & 0.28$\pm$0.01	\\
210418 & A1 V        & 3.69 & 3.60 & 3.52 & 3.47 & 3.44 & 3.46 & 3.39 & 3.38 & 8511  & 4.02 & 2 & 0.03 & 7  & 0.78$\pm$0.04	\\
211336 & F0 V        & 4.51 & 4.47 & 4.19 & 4.03 & 3.89 & 3.84 & 3.67 & 3.54 & 7586  & 4.13 & 2 & 0.03 & 11 & 0.69$\pm$0.03	\\
212061 & A0 V        & 3.68 & 3.79 & 3.85 & 3.86 & 3.91 & 4.11 & 4.05 & 4.02 & 10471 & 4.11 & 2 & 0.02 & 7  & 0.52$\pm$0.03	\\
213558 & A1 V        & 3.78 & 3.77 & 3.76 & 3.75 & 3.76 & 3.83 & 3.87 & 3.85 & 9333  & 4.20 & 2 & 0.00 & 10 & 0.60$\pm$0.03	\\
214923 & B8 V        & 3.10 & 3.32 & 3.41 & 3.43 & 3.51 & 3.54 & 3.53 & 3.57 & 10965 & 3.75 & 2 & 0.01 & 7  & 0.60$\pm$0.03	\\
219688 & B7/8 V      & 3.69 & 4.24 & 4.39 & 4.46 & 4.60 & 4.70 & 4.76 & 4.76 & 14125 & 4.13 & 2 & 0.02 & 9  & 0.30$\pm$0.02	\\
224617 & F4 V        & 4.47 & 4.43 & 4.01 & 3.79 & 3.59 & 3.51 & 3.30 & 3.10 & 6761  & 3.66 & 2 & 0.01 & 15 & 0.92$\pm$0.05	\\
\enddata
\tablecomments{Spectral types are from SIMBAD; $UBV$ values are from \citet{Mermilliod}; $RI$ values are from \citet{2003AJ....125..984M}; $JHK$ values are from \citet{2003yCat.2246....0C}; $T_{\rm eff}$, log $g$, and $E(B-V)$ values are from the following sources: 1: \citet{2007astro.ph..3658P}; 2. \citet{1999AandA...352..555A}; 3. \citet{2011AandA...525A..71W}; 4. \citet{2010SPIE.7734E..4EL}; 5. \citet{2006MNRAS.371..703S}; 6. \citet{1980BICDS..19...61N}; 7. \citet{2009AandA...501..297Z}; 8. \citet{1995yCat.3039....0J}; 9. \citet{1992BICDS..40...31F}; 10. \citet{1996AandAS..117..227A}; 11. \citet{2018yCat.2354....0G}; 12. \citet{2008ApJS..176..276V}; 13. \citet{2006ApJ...647..312O}; 14. \citet{2012ApJS..199....8G}; and 15. \citet{2006MNRAS.371.1793K}. $\theta_{\rm est}$ is the estimated angular diameter calculated using the method described in Section 2.}
\end{deluxetable}
%\end{longrotatetable}

%%%%%%%%%%%%%%%%%%%%%% Interferometric Results %%%%%%%%%%%%%%%%%%%%%%%%%%%%%%%%%%%%%%%%%%%%%%%

%\startlongtable
%\begin{longrotatetable}
\begin{deluxetable}{cccccccccccc}
%\rotate
\tablewidth{0pc}
\tabletypesize{\scriptsize}
\tablecaption{Interferometric Results. \label{inf_results}}

\tablehead{\colhead{Target} & \colhead{$\theta_{\rm UD}$} & \colhead{$T_{\rm eff}$} & \colhead{log $g$}        & \colhead{}    & \colhead{Initial}             & \colhead{$\theta_{\rm LD,initial}$} & \colhead{Final}               & \colhead{$\theta_{\rm LD,final}$} & \colhead{$\sigma_{\rm LD}$} & \colhead{Max SF} & \colhead{$\#$} \\ 
           \colhead{HD}     & \colhead{(mas)}             & \colhead{(K)}           & \colhead{(cm s$^{-2}$)}  & \colhead{Ref} & \colhead{$\mu_{\rm \lambda}$} & \colhead{(mas)}             & \colhead{$\mu_{\rm \lambda}$} & \colhead{(mas)}             & \colhead{($\%$)} & \colhead{(10$^6$ cycles s$^{\rm -1}$)}   & \colhead{pts}            }
\startdata
3627   & 3.879$\pm$0.040  & 4365  & 2.38  & 1 & 0.75 & 4.203$\pm$0.036  & 0.72 & 4.185$\pm$0.036  & 0.9 & 94.9  & 165  \\
4656   & 3.559$\pm$0.042  & 3956  & 1.85  & 2 & 0.77 & 3.814$\pm$0.035  & 0.81 & 3.841$\pm$0.035  & 0.9 & 81.2  & 531  \\
5112   & 3.683$\pm$0.082  & 3690  & 1.63  & 3 & 0.81 & 3.730$\pm$0.041  & 0.81 & 3.730$\pm$0.041  & 1.1 & 114.5 & 477  \\
6805   & 3.155$\pm$0.014  & 4467  & 2.29  & 1 & 0.71 & 3.304$\pm$0.012  & 0.71 & 3.304$\pm$0.012  & 0.4 & 139.5 & 2986 \\
9826   & 1.046$\pm$0.017  & 6457  & 4.26  & 1 & 0.51 & 1.080$\pm$0.018  & 0.54 & 1.083$\pm$0.018  & 1.7 & 94.8  & 4209 \\
10380  & 2.752$\pm$0.041  & 4170  & 1.91  & 2 & 0.73 & 2.885$\pm$0.045  & 0.70 & 2.873$\pm$0.045  & 1.6 & 63.4  & 405  \\
19373  & 0.987$\pm$0.031  & 6166  & 4.33  & 1 & 0.54 & 1.020$\pm$0.031  & 0.51 & 1.017$\pm$0.031  & 3.0 & 67.2  & 2966 \\
20902  & 3.066$\pm$0.015  & 6690  & 1.31  & 2 & 0.55 & 3.180$\pm$0.008  & 0.55 & 3.180$\pm$0.008  & 0.3 & 71.1  & 2360 \\
25025  & 8.560$\pm$0.023  & 3643  & 1.52  & 3 & 0.81 & 9.286$\pm$0.028  & 0.81 & 9.286$\pm$0.028  & 0.3 & 28.1  & 478  \\
28307  & 2.089$\pm$0.040  & 4898  & 2.59  & 1 & 0.67 & 2.178$\pm$0.033  & 0.65 & 2.172$\pm$0.033  & 1.5 & 93.5  & 685  \\
35497  & 1.180$\pm$0.069  & 12589 & 3.63  & 1 & 0.34 & 1.238$\pm$0.052  & 0.35 & 1.239$\pm$0.052  & 4.2 & 67.8  & 719  \\
39003  & 2.544$\pm$0.022  & 4571  & 2.18  & 1 & 0.71 & 2.681$\pm$0.027  & 0.71 & 2.681$\pm$0.027  & 1.0 & 39.5  & 5921 \\
42995  & 11.072$\pm$0.026 & 3600  & 1.50  & 4 & 0.81 & 12.112$\pm$0.024 & 0.81 & 12.112$\pm$0.024 & 0.2 & 24.5  & 500  \\
60522  & 4.551$\pm$0.026  & 3846  & 1.69  & 2 & 0.78 & 4.765$\pm$0.030  & 0.76 & 4.748$\pm$0.030  & 0.6 & 89.5  & 1113 \\
61421  & 5.291$\pm$0.012  & 6761  & 4.04  & 1 & 0.49 & 5.406$\pm$0.006  & 0.49 & 5.406$\pm$0.006  & 0.1 & 94.8  & 8812 \\
61935  & 2.130$\pm$0.023  & 4786  & 2.47  & 1 & 0.68 & 2.179$\pm$0.021  & 0.65 & 2.170$\pm$0.021  & 1.0 & 139.4	& 911  \\
74442  & 2.423$\pm$0.024  & 4677  & 2.59  & 1 & 0.69 & 2.595$\pm$0.021  & 0.69 & 2.595$\pm$0.021  & 0.8 & 138.0	& 5090 \\
80493  & 7.315$\pm$0.038  & 3873  & 1.78  & 3 & 0.79 & 7.930$\pm$0.027  & 0.81 & 7.954$\pm$0.027  & 0.3 & 92.8	& 1205 \\
89758  & 7.938$\pm$0.040  & 3700  & 1.35  & 4 & 0.81 & 8.592$\pm$0.029  & 0.80 & 8.579$\pm$0.029  & 0.3 & 27.4  & 880  \\
102224 & 3.175$\pm$0.038  & 4467  & 1.97  & 1 & 0.69 & 3.530$\pm$0.022  & 0.71 & 3.541$\pm$0.022  & 0.6 & 140.7 & 2280 \\
109387 & 0.909$\pm$0.042  & 14859 & 3.15  & 4 & 0.33 & 0.898$\pm$0.048  & 0.43 & 0.906$\pm$0.049  & 5.4 & 60.6	& 4593 \\
112300 & 10.041$\pm$0.025 & 3652  & 1.30  & 4 & 0.82 & 10.918$\pm$0.021 & 0.82 & 10.918$\pm$0.021 & 0.2 & 28.5  & 1360 \\
129989 & 4.588$\pm$0.011  & 4730  & 2.24  & 4 & 0.68 & 4.840$\pm$0.010  & 0.68 & 4.840$\pm$0.010  & 0.2 & 81.2	& 1915 \\
131873 & 9.484$\pm$0.009  & 4030  & 1.83  & 5 & 0.77 & 10.229$\pm$0.012 & 0.77 & 10.229$\pm$0.012 & 0.1 & 28.5	& 1708 \\
132813 & 9.507$\pm$0.016  & 3457  & 0.14  & 6 & 0.86 & 10.442$\pm$0.021 & 0.86 & 10.442$\pm$0.021 & 0.2 & 34.3  & 1389 \\
133165 & 2.033$\pm$0.038  & 4786  & 2.42  & 1 & 0.67 & 2.140$\pm$0.014  & 0.69 & 2.147$\pm$0.014  & 0.7 & 150.7	& 1076 \\
135722 & 2.685$\pm$0.027  & 4898  & 2.48  & 1 & 0.65 & 2.870$\pm$0.012  & 0.67 & 2.878$\pm$0.012  & 0.4 & 110.5	& 883  \\
141477 & 5.229$\pm$0.013  & 3643  & 1.52  & 3 & 0.81 & 5.663$\pm$0.021  & 0.80 & 5.653$\pm$0.021  & 0.4 & 61.9  & 986  \\
163993 & 2.104$\pm$0.016  & 4898  & 2.64  & 1 & 0.67 & 2.213$\pm$0.017  & 0.65 & 2.206$\pm$0.017  & 0.8 & 132.3 & 527  \\
164058 & 9.428$\pm$0.013  & 3990  & 1.64  & 2 & 0.77 & 10.190$\pm$0.015 & 0.77 & 10.190$\pm$0.015 & 0.1 & 28.5  & 1172 \\
175588 & 10.487$\pm$0.017 & 3382  & 0.55  & 2 & 0.84 & 11.541$\pm$0.024 & 0.84 & 11.541$\pm$0.024 & 0.2 & 24.5  & 684  \\
189319 & 5.743$\pm$0.014  & 3690  & 1.63  & 3 & 0.81 & 6.131$\pm$0.011  & 0.77 & 6.089$\pm$0.011  & 0.2 & 62.1  & 3960 \\
192909 & 5.137$\pm$0.033  & 3978  & 1.03  & 2 & 0.76 & 5.514$\pm$0.015  & 0.81 & 5.557$\pm$0.015  & 0.3 & 39.9  & 3001 \\
197989 & 4.640$\pm$0.062  & 4786  & 2.67  & 1 & 0.68 & 4.978$\pm$0.046  & 0.69 & 4.985$\pm$0.046  & 0.9 & 33.5  & 350  \\
198026 & 6.472$\pm$0.075  & 3627  & 0.25  & 6 & 0.86 & 7.124$\pm$0.088  & 0.82 & 7.079$\pm$0.088  & 1.2 & 23.8  & 140  \\
200905 & 5.450$\pm$0.013  & 4031  & 0.89  & 2 & 0.77 & 5.797$\pm$0.010  & 0.79 & 5.816$\pm$0.010  & 0.2 & 62.1  & 4230 \\
208816 & 6.767$\pm$0.018  & 3465  & -0.22 & 6 & 0.85 & 7.238$\pm$0.012  & 0.86 & 7.251$\pm$0.012  & 0.2 & 49.8  & 4394 \\
210745 & 5.003$\pm$0.022  & 4340  & 0.88  & 2 & 0.74 & 5.285$\pm$0.023  & 0.76 & 5.302$\pm$0.023  & 0.4 & 67.8  & 766  \\
213306 & 1.455$\pm$0.013  & 5864  & 1.65  & 7 & 0.54 & 1.512$\pm$0.018  & 0.62 & 1.526$\pm$0.018  & 1.2 & 71.1  & 4124 \\
214868 & 2.406$\pm$0.024  & 4318  & 2.50  & 3 & 0.74 & 2.569$\pm$0.047  & 0.70 & 2.555$\pm$0.047  & 1.8 & 63.4  & 361  \\
216131 & 2.338$\pm$0.006  & 5012  & 2.60  & 1 & 0.65 & 2.508$\pm$0.125  & 0.65 & 2.508$\pm$0.125  & 0.5 & 113.6 & 4650 \\ 
216386 & 7.742$\pm$0.029  & 3702  & 0.49  & 6 & 0.82 & 8.473$\pm$0.042  & 0.71 & 8.333$\pm$0.042  & 0.5 & 24.2  & 380  \\ 
218452 & 1.858$\pm$0.023  & 4046  & 1.93  & 3 & 0.78 & 2.003$\pm$0.047  & 0.74 & 1.991$\pm$0.047  & 2.4 & 121.0 & 160  \\
224935 & 7.297$\pm$0.052  & 3551  & 0.37  & 6 & 0.85 & 8.007$\pm$0.059  & 0.85 & 8.007$\pm$0.059  & 0.7 & 24.2	& 230  \\
\enddata 
\tablecomments{The initial $\mu_{\lambda}$ is based on the $T_{\rm eff}$ and log $g$ listed in the table, and the final $\mu_{\lambda}$ is based on the new $T_{\rm eff}$ determination. (See Section 3.2 for more details). The $T_{\rm eff}$ and log $g$ are from the following sources: 1. \citet{1999AandA...352..555A}; 2. \citet{2011AandA...531A.165P}; 3. \citet{2002AandA...393..183B}; 4. \citet{2016AandA...591A.118S}; 5. \citet{2004ApJS..152..251V}; 6. \citet{2017MNRAS.471..770M}; and 7. \citet{2013AandA...549A.129C}. 
Max SF is the maximum spatial frequency for that star's diameter measurement. $\#$ pts is the number of data points in the angular diameter fit.}
\end{deluxetable}
%\end{longrotatetable}

\clearpage

%%%%%%%%%%%%%%%%%%%%%% Derived Results %%%%%%%%%%%%%%%%%%%%%%%%%%%%%%%%%%%%%%%%%%%%%%%

%\startlongtable 
%\begin{longrotatetable} 
\begin{deluxetable}{ccccccccc}
%\rotate
\tablewidth{0pc}
\tabletypesize{\scriptsize}
\tablecaption{Derived Stellar Parameters. \label{derived_results}}

\tablehead{\colhead{Target} & \colhead{Spectral}       & \colhead{$R$}         & \colhead{$\sigma_R$} & \colhead{$F_{\rm BOL}$}                      & \colhead{$A_{\rm V}$}           & \colhead{$T_{\rm eff}$} & \colhead{$\sigma_T$} & \colhead{$L$}        \\ 
           \colhead{HD}     & \colhead{Type}           & \colhead{($R_\odot$)} & \colhead{($\%$)}     & \colhead{(10$^{-6}$ erg s$^{-1}$ cm$^{-2}$)} & \colhead{(mag)}                 & \colhead{(K)}           & \colhead{($\%$)}        & \colhead{($L_\odot$)} }
\startdata
3627   & K2 III & 14.25$\pm$$^{\rm 0.23}_{\rm 0.24}$    & 1.7  & 2.35$\pm$0.09  & 0.20$\pm$0.05 & 4480$\pm$48   & 1.1 & 73.8$\pm$3.6	    \\
4656   & K5 III & 38.01$\pm$$^{\rm 0.71}_{\rm 0.73}$    & 1.9  & 1.10$\pm$0.03  & 0.00$\pm$0.05 & 3868$\pm$35   & 0.9 & 291.7$\pm$13.3	\\
5112   & K5 III & 65.93$\pm$$^{\rm 2.13}_{\rm 2.25}$    & 3.4  & 0.89$\pm$0.03  & 0.05$\pm$0.05 & 3724$\pm$35   & 0.9 & 754.8$\pm$52.6	\\
6805   & G8 III & 13.12$\pm$0.10                        & 0.8  & 1.99$\pm$0.04  & 0.44$\pm$0.04 & 4836$\pm$23   & 0.5 & 85.0$\pm$1.9	    \\
9826   & G4 III & 1.57$\pm$0.03                         & 1.7  & 0.55$\pm$0.02  & 0.00$\pm$0.06 & 6114$\pm$77   & 1.3 & 3.1$\pm$0.1	    \\
10380  & K2 III & 34.58$\pm$$^{\rm 0.81}_{\rm 0.83}$    & 2.4  & 1.07$\pm$0.04  & 0.49$\pm$0.05 & 4441$\pm$58   & 1.3 & 419.7$\pm$23.1	\\
19373  & G4 III & 1.15$\pm$0.04                         & 3.1  & 0.60$\pm$0.02  & 0.00$\pm$0.07 & 6449$\pm$117  & 1.8 & 2.1$\pm$0.1	    \\
20902  & G4 III & 53.07$\pm$$^{\rm 1.37}_{\rm 1.45}$    & 2.7  & 3.97$\pm$0.11  & 0.00$\pm$0.06 & 5859$\pm$41   & 0.7 & 2994$\pm$178	\\
25025  & K5 III & 58.70$\pm$$^{\rm 0.80}_{\rm 0.82}$    & 1.4  & 5.86$\pm$0.21  & 0.28$\pm$0.05 & 3779$\pm$34   & 0.9 & 634.2$\pm$28.6	\\
28307  & G7 III & 9.42$\pm$0.22                         & 2.4  & 0.94$\pm$0.03  & 0.00$\pm$0.04 & 4940$\pm$55   & 1.1 & 47.7$\pm$2.3	    \\
35497  & B7 III & 5.47$\pm$0.24                         & 4.4  & 10.70$\pm$0.25 & 0.00$\pm$0.03 & 12026$\pm$262 & 2.2 & 564.0$\pm$20.7	\\
39003  & G9 III & 17.85$\pm$$^{\rm 0.51}_{\rm 0.53}$    & 3.0  & 1.05$\pm$0.04  & 0.20$\pm$0.04 & 4576$\pm$50   & 1.1 & 126.1$\pm$8.4	\\
42995  & M3 III & 275.20$\pm$$^{\rm 48.82}_{\rm 75.66}$ & 27.5 & 7.35$\pm$0.25  & 0.09$\pm$0.05 & 3502$\pm$30   & 0.8 & 10276$\pm$4445	\\
60522  & K5 III & 39.62$\pm$$^{\rm 0.74 }_{\rm 0.76}$   & 1.9  & 1.96$\pm$0.07  & 0.22$\pm$0.05 & 4019$\pm$38   & 0.9 & 369.6$\pm$18.5	\\
61421  & F5 IV  & 2.04$\pm$0.01                         & 0.5  & 17.90$\pm$0.62 & 0.00$\pm$0.05 & 6548$\pm$57   & 0.9 & 6.9$\pm$0.3	    \\
61935  & G6 III & 10.42$\pm$0.12                        & 1.1  & 1.02$\pm$0.02  & 0.23$\pm$0.04 & 5049$\pm$32   & 0.6 & 63.7$\pm$1.3	    \\
74442  & G9 III & 11.70$\pm$0.13                        & 1.1  & 1.08$\pm$0.02  & 0.26$\pm$0.04 & 4684$\pm$27   & 0.6 & 59.5$\pm$1.4	    \\
80493  & K5 III & 58.15$\pm$$^{\rm 0.73}_{\rm 0.75}$    & 1.3  & 4.35$\pm$0.15  & 0.17$\pm$0.05 & 3790$\pm$33   & 0.9 & 629.7$\pm$26.3	\\
89758  & K5 III & 51.80$\pm$$^{\rm 1.12}_{\rm 1.17}$    & 2.3  & 5.49$\pm$0.21  & 0.34$\pm$0.05 & 3868$\pm$37   & 0.9 & 542.0$\pm$31.2	\\
102224 & K2 III & 23.15$\pm$0.21                        & 0.9  & 1.47$\pm$0.04  & 0.16$\pm$0.03 & 4331$\pm$33   & 0.8 & 170.1$\pm$5.3	\\
109387 & B6 III & 13.89$\pm$$^{\rm 0.94}_{\rm 0.97}$    & 7.0  & 1.88$\pm$0.05  & 0.11$\pm$0.03 & 9105$\pm$254  & 2.8 & 1197$\pm$107	\\
112300 & M3 III & 67.40$\pm$$^{\rm 0.96}_{\rm 0.99}$    & 1.5  & 6.30$\pm$0.20  & 0.06$\pm$0.05 & 3549$\pm$28   & 0.8 & 650.1$\pm$27.5	\\
129989 & K0 III & 37.61$\pm$$^{\rm 1.29}_{\rm 1.38}$    & 3.7  & 3.99$\pm$0.22  & 0.11$\pm$0.06 & 4755$\pm$66   & 1.4 & 652.5$\pm$58.7	\\
131873 & K4 III & 44.13$\pm$0.22                        & 0.5  & 9.00$\pm$0.33  & 0.09$\pm$0.05 & 4008$\pm$37   & 0.9 & 453.7$\pm$17.2	\\
132813 & M4 III & 113.93$\pm$$^{\rm 5.40}_{\rm 5.97}$   & 5.2  & 4.91$\pm$0.21  & 0.59$\pm$0.05 & 3410$\pm$37   & 1.1 & 1583$\pm$172	\\
133165 & G8 III & 13.78$\pm$0.14                        & 1.0  & 0.72$\pm$0.01  & 0.24$\pm$0.04 & 4655$\pm$24   & 0.5 & 80.4$\pm$1.8	    \\
135722 & K0 III & 11.43$\pm$0.07                        & 0.6  & 1.35$\pm$0.04  & 0.01$\pm$0.05 & 4703$\pm$38   & 0.8 & 57.6$\pm$1.9	    \\
141477 & K6 III & 71.31$\pm$$^{\rm 1.42}_{\rm 1.48}$    & 2.1  & 2.37$\pm$0.09  & 0.41$\pm$0.05 & 3863$\pm$39   & 1.0 & 1021$\pm$57	\\
163993 & G7 III & 9.94$\pm$0.09                         & 0.9  & 1.04$\pm$0.04  & 0.00$\pm$0.06 & 5032$\pm$48   & 1.0 & 57.2$\pm$2.1	    \\
164058 & K5 III & 51.80$\pm$0.26                        & 0.5  & 8.54$\pm$0.29  & 0.00$\pm$0.04 & 3964$\pm$34   & 0.9 & 597.7$\pm$21.2	\\
175588 & M4 III & 292.53$\pm$$^{\rm 19.34}_{\rm 22.28}$ & 7.6  & 5.89$\pm$0.22  & 0.37$\pm$0.05 & 3394$\pm$32   & 0.9 & 10248$\pm$1500	\\
189319 & K5 III & 57.71$\pm$$^{\rm 0.86}_{\rm 0.88}$    & 1.5  & 3.02$\pm$0.12  & 0.15$\pm$0.06 & 3954$\pm$38   & 1.0 & 734.6$\pm$35.8	\\
192909 & K7 III & 183.20$\pm$$^{\rm 10.10}_{\rm 11.35}$ & 6.2  & 1.94$\pm$0.07  & 0.10$\pm$0.06 & 3706$\pm$35   & 0.9 & 5710$\pm$699	\\
197989 & K0 III & 12.41$\pm$$^{\rm 0.29}_{\rm 0.30}$    & 2.4  & 3.90$\pm$0.09  & 0.15$\pm$0.02 & 4659$\pm$35   & 0.7 & 65.4$\pm$3.2	    \\
198026 & M3 III & 117.59$\pm$$^{\rm 3.98}_{\rm 4.21}$   & 3.6  &  2.37$\pm$0.08 & 0.08$\pm$0.05 & 3452$\pm$35   & 1.0 & 1771$\pm$128	\\
200905 & K5 III & 220.09$\pm$$^{\rm 9.64}_{\rm 10.56}$  & 4.8  & 2.55$\pm$0.09  & 0.16$\pm$0.05 & 3878$\pm$33   & 0.8 & 9889$\pm$964	\\
208816 & M3 III & 779.27$\pm$$^{\rm 77.24}_{\rm 96.32}$ & 12.4 & 2.33$\pm$0.10  & 0.51$\pm$0.05 & 3396$\pm$35   & 1.0 & 72881$\pm$16307	\\
210745 & K1 III & 172.67$\pm$$^{\rm 7.54}_{\rm 8.26}$   & 4.8  & 3.49$\pm$0.18  & 0.81$\pm$0.05 & 4393$\pm$58   & 1.3 & 10024$\pm$1052	\\
213306 & G4 III & 46.07$\pm$$^{\rm 1.94}_{\rm 2.10}$    & 4.6  & 0.60$\pm$0.02  & 0.00$\pm$0.05 & 5273$\pm$50   & 0.9 & 1481$\pm$132	\\
214868 & K2 III & 29.46$\pm$$^{\rm 0.64}_{\rm 0.65}$    & 2.2  & 0.89$\pm$0.04  & 0.32$\pm$0.06 & 4494$\pm$64   & 1.4 & 319.4$\pm$15.7	\\
216131 & G7 III & 9.32$\pm$0.47                         & 5.0  & 1.28$\pm$0.05  & 0.01$\pm$0.06 & 4971$\pm$132  & 2.7 & 47.8$\pm$1.8	    \\
216386 & G8 III & 100.17$\pm$$^{\rm 2.67}_{\rm 2.81}$   & 2.8  & 10.50$\pm$2.42 & 2.47$\pm$0.05 & 4616$\pm$266  & 5.8 & 4109$\pm$973	\\
218452 & K2 III & 23.36$\pm$0.59                        & 2.5  & 0.47$\pm$0.12  & 0.43$\pm$0.19 & 4332$\pm$291  & 6.7 & 173.3$\pm$46.0	\\
224935 & M3 III & 109.20$\pm$$^{\rm 5.46}_{\rm 6.05}$   & 5.5  & 3.17$\pm$0.12  & 0.36$\pm$0.05 & 3490$\pm$35   & 1.0 & 1597$\pm$177	\\
\enddata
\tablecomments{The spectral types are those that provide the best SED fit as described in Section 3.2. The SED fits are also the source of $F_{\rm BOL}$ and $A_{\rm V}$. The other parameters are derived as described in Section 3.2.}
\end{deluxetable}
%\end{longrotatetable}

\clearpage

%%%%%%%%%%%%%%%%%%%%%% Literature and SED Comparison %%%%%%%%%%%%%%%%%%%%%%%%%%%%%%%%%%%%%%%%%%%%%%%

\startlongtable
\begin{deluxetable*}{cccl}
%\rotate
%\tablewidth{0pc}
\tabletypesize{\scriptsize}
\tablecaption{Interferometric Angular Diameter Comparison. \label{lit}}

\tablehead{\colhead{Target} & \colhead{$\theta_{\rm LD,here}$} & \colhead{$\theta_{\rm LD,literature}$} & \colhead{ }      \\ 
           \colhead{HD}     & \colhead{(mas)}                  & \colhead{(mas)}                      & \colhead{Reference} } 
\startdata
3627   & 4.185$\pm$0.036  &   4.12$\pm$0.04         & \citet{1991AJ....101.2207M}	    \\	
       &                  &   4.24$\pm$0.06         & \citet{2001AJ....122.2707N}	    \\	% NPOI      
       &                  &   4.17$\pm$0.06         & \citet{2001AJ....122.2707N}	    \\	% Mark III
       &                  &  4.136$\pm$0.041        & \citet{2003AJ....126.2502M}	    \\	
4656   & 3.841$\pm$0.035  &   3.80$\pm$0.10$^\ast$  & \citet{2005AandA...434.1201R}	    \\	% UD
5112   & 3.730$\pm$0.041  &   3.40$\pm$0.04$^\ast$  & \citet{2005AandA...434.1201R}	    \\	% UD
       &                  &  3.459$\pm$0.006        & \citet{2009MNRAS.399..399R}	    \\	
       &                  &  3.375$\pm$0.016        & \citet{2019MNRAS.490.3158C}	    \\	
6805   & 3.304$\pm$0.012  &   3.35$\pm$0.04$^\ast$  & \citet{2005AandA...434.1201R} 	\\	% UD
       &                  &  3.235$\pm$0.016        & \citet{2019MNRAS.490.3158C}	    \\	
9826   & 1.083$\pm$0.018  &  1.114$\pm$0.009        & \citet{2008ApJ...680..728B}	    \\	
       &                  &  1.161$\pm$0.027        & \citet{2016AandA...586A..94L} 	\\	
10380  & 2.873$\pm$0.045  &   2.81$\pm$0.03         & \citet{1999AJ....118.3032N}	    \\	
       &                  &   2.93$\pm$0.13$^\ast$  & \citet{2005AandA...434.1201R}	    \\	% UD
       &                  &  2.902$\pm$0.013        & \citet{2019MNRAS.490.3158C}	    \\	
19373  & 1.017$\pm$0.031  &  1.331$\pm$0.050        & \citet{2009ApJ...694.1085V}	    \\	
       &                  &  1.246$\pm$0.008        & \citet{2012ApJ...746..101B}	    \\	
20902  & 3.180$\pm$0.008  &   3.10$\pm$0.02         & \citet{1999AJ....118.3032N}	    \\	
       &                  &   3.12$\pm$0.03         & \citet{2001AJ....122.2707N}	    \\	% NPOI      
       &                  &   3.23$\pm$0.05         & \citet{2001AJ....122.2707N}	    \\	% Mark III
       &                  &  3.188$\pm$0.035        & \citet{2003AJ....126.2502M}	    \\	
25025  & 9.286$\pm$0.028  &  9.332$\pm$0.173        & \citet{2003AJ....126.2502M}	    \\	
       &                  &   8.48$\pm$0.09         & \citet{2005AandA...434.1201R}	    \\	
       &                  &  8.908$\pm$0.461        & \citet{2009MNRAS.399..399R}	    \\	
28307  & 2.172$\pm$0.033  &  2.305$\pm$0.043        & \citet{2009ApJ...691.1243B}	    \\	
       &                  &  2.169$\pm$0.019        & \citet{2019MNRAS.490.3158C}	    \\	
35497  & 1.239$\pm$0.052  &   1.56$\pm$0.11$^\ast$  & \citet{1999AJ....117..521V}	    \\	% UD
       &                  &  1.090$\pm$0.076        & \citet{2019ApJ...873...91G}	    \\	
39003  & 2.681$\pm$0.027  &   2.73$\pm$0.06$^\ast$  & \citet{1999AJ....117..521V}	    \\	% UD
42995  & 12.112$\pm$0.024 &  11.75$\pm$0.27$^\ast$  & \citet{1993ApJ...406..215Q}	    \\	% UD
       &                  & 11.789$\pm$0.118        & \citet{2003AJ....126.2502M}	    \\	
       &                  &  12.57$\pm$0.04$^\ast$  & \citet{2005AandA...434.1201R}	    \\	% UD
60522  & 4.748$\pm$0.030  &  4.789$\pm$0.021        & \citet{2019MNRAS.490.3158C}	    \\	
61421  & 5.406$\pm$0.006  &   5.76$\pm$0.10         & \citet{1967MNRAS.137..393H}	    \\	
       &                  &   5.50$\pm$0.17         & \citet{1974MNRAS.167..121H}	    \\	
       &                  &   6.44$\pm$0.25         & \citet{1988ApJ...327..905S}	    \\	
       &                  &   5.51$\pm$0.05         & \citet{1991AJ....101.2207M}	    \\	
       &                  &   5.43$\pm$0.07         & \citet{2001AJ....122.2707N}	    \\	% NPOI      
       &                  &   5.46$\pm$0.08         & \citet{2001AJ....122.2707N}	    \\	% Mark III
       &                  &  5.446$\pm$0.054        & \citet{2003AJ....126.2502M}	    \\	
       &                  &  5.448$\pm$0.053        & \citet{2004AandA...413..251K}	    \\	
       &                  &   5.37$\pm$0.11         & \citet{2005AandA...434.1201R}	    \\	
       &                  &  5.368$\pm$0.078        & \citet{2009MNRAS.399..399R}	    \\	
61935  & 2.170$\pm$0.021  &  2.243$\pm$0.009        & \citet{2019MNRAS.490.3158C}	    \\	
74442  & 2.595$\pm$0.021  &  2.389$\pm$0.012        & \citet{2019MNRAS.490.3158C}	    \\	
80493  & 7.954$\pm$0.027  &   7.98$\pm$0.31         & \citet{1987AandA...188..114D}	    \\	
       &                  &   7.50$\pm$0.09         & \citet{2001AJ....122.2707N}	    \\	% NPOI      
       &                  &   7.59$\pm$0.11         & \citet{2001AJ....122.2707N}	    \\	% Mark III
       &                  &  7.538$\pm$0.075        & \citet{2003AJ....126.2502M}	    \\	
       &                  &  7.145$\pm$0.064        & \citet{2019MNRAS.490.3158C}	    \\	
89758  & 8.579$\pm$0.029  &   8.69$\pm$0.09         & \citet{2001AJ....122.2707N}	    \\	% NPOI      
       &                  &   8.55$\pm$0.12         & \citet{2001AJ....122.2707N}	    \\	% Mark III
       &                  &  8.538$\pm$0.085        & \citet{2003AJ....126.2502M}	    \\	
102224 & 3.541$\pm$0.022  &   3.23$\pm$0.02         & \citet{1999AJ....118.3032N}	    \\	
109387 & 0.906$\pm$0.049  &       N/A               & 	                                \\	
112300 & 10.918$\pm$0.021 &    9.8$\pm$0.6$^\ast$   & \citet{1998AJ....116..981D}	    \\	% UD
       &                  & 10.709$\pm$0.107        & \citet{2003AJ....126.2502M}	    \\	
129989 & 4.840$\pm$0.010  &       N/A               &                  	                \\	
131873 & 10.229$\pm$0.012 &    8.9$\pm$1.1$^\ast$   & \citet{1983AandA...120..263F}	    \\	% UD
       &                  &    9.7$\pm$0.8$^\ast$   & \citet{1998AJ....116..981D}       \\	% UD
       &                  & 10.301$\pm$0.103        & \citet{2003AJ....126.2502M}	    \\	
132813 & 10.442$\pm$0.021 &    9.6$\pm$0.7$^\ast$   & \citet{1998AJ....116..981D}	    \\	% UD
       &                  & 10.588$\pm$0.170        & \citet{2003AJ....126.2502M}	    \\	
133165 & 2.147$\pm$0.014  &  1.934$\pm$0.008        & \citet{2019MNRAS.490.3158C}	    \\	
135722 & 2.878$\pm$0.012  &   2.74$\pm$0.03         & \citet{1999AJ....118.3032N}	    \\	
       &                  &   2.65$\pm$0.06$^\ast$  & \citet{1999AJ....117..521V}	    \\	% UD
       &                  &   2.76$\pm$0.03         & \citet{2001AJ....122.2707N}	    \\	% NPOI      
       &                  &   2.75$\pm$0.04         & \citet{2001AJ....122.2707N}	    \\	% Mark III
       &                  &  2.764$\pm$0.030        & \citet{2003AJ....126.2502M}	    \\	
141477 & 5.653$\pm$0.021  &    6.2$\pm$0.5$^\ast$   & \citet{1998AJ....116..981D}	    \\	% UD
       &                1  &  5.34$\pm$0.06$^\ast$  & \citet{2005AandA...434.1201R}	    \\	% UD
163993 & 2.206$\pm$0.017  &  2.196$\pm$0.010        & \citet{2019MNRAS.490.3158C}	    \\	
164058 & 10.190$\pm$0.015 &   10.2$\pm$1.4          & \citet{1983ApJ...268..309D}	    \\	
       &                  &    7.9$\pm$0.7$^\ast$   & \citet{1983AandA...120..263F}	    \\	% UD
       &                  &  10.13$\pm$0.24         & \citet{1987AandA...188..114D}	    \\	
       &                  &    9.6$\pm$0.3$^\ast$   & \citet{1996AJ....111.1705D}	    \\	% UD
       &                  &    9.6$\pm$0.3$^\ast$   & \citet{1998AJ....116..981D}	    \\	% UD
       &                  &  9.860$\pm$0.128        & \citet{2003AJ....126.2502M}	    \\	
175588 & 11.541$\pm$0.024 &  11.76$\pm$0.30$^\ast$  & \citet{1993ApJ...406..215Q}	    \\	% UD
       &                  &    9.6$\pm$0.4$^\ast$   & \citet{1996AJ....111.1705D}	    \\	% UD
       &                  &    9.7$\pm$0.3$^\ast$   & \citet{1998AJ....116..981D}	    \\	% UD
       &                  &  11.40$\pm$0.43         & \citet{2002AJ....124.3370S}	    \\	
       &                  & 11.530$\pm$0.156        & \citet{2003AJ....126.2502M}	    \\	
189319 & 6.089$\pm$0.011  &    4.6$\pm$1.1$^\ast$   & \citet{1996AJ....111.1705D}	    \\	% UD
       &                  &    5.5$\pm$0.5$^\ast$  & \citet{1998AJ....116..981D}	    \\	% UD
       &                  &   6.18$\pm$0.07         & \citet{2001AandA...377..981W} 	\\	
       &                  &  6.225$\pm$0.062        & \citet{2003AJ....126.2502M}	    \\	
       &                  &   5.83$\pm$0.41$^\ast$  & \citet{2005AandA...434.1201R}	    \\	% UD
       &                  &  6.184$\pm$0.048        & \citet{2009MNRAS.399..399R}	    \\	
       &                  &  5.935$\pm$0.027        & \citet{2019MNRAS.490.3158C}	    \\	
192909 & 5.557$\pm$0.015  &    5.5$\pm$0.5          & \citet{1990AandA...236..449D} 	\\	
       &                  &    6.2$\pm$0.6$^\ast$   & \citet{1996AJ....111.1705D}	    \\	% UD
       &                  &    6.2$\pm$0.6$^\ast$   & \citet{1998AJ....116..981D}	    \\	% UD
       &                  &   5.16$\pm$0.12         & \citet{2001AJ....122.2707N}   	\\	% NPOI      
       &                  &   5.46$\pm$0.08         & \citet{2001AJ....122.2707N}   	\\	% Mark III
       &                  &  5.423$\pm$0.054        & \citet{2003AJ....126.2502M}	    \\	                                    
197989 & 4.985$\pm$0.046  &   4.62$\pm$0.040        & \citet{1991AJ....101.2207M}   	\\	
       &                  &  4.612$\pm$0.046        & \citet{2003AJ....126.2502M}	    \\	
       &                  &   4.61$\pm$0.02         & \citet{2017AandA...600L...2C} 	\\	
       &                  &  4.564$\pm$0.016        & \citet{2019MNRAS.490.3158C}	    \\	
198026 & 7.079$\pm$0.088  &    5.5$\pm$0.7$^\ast$   & \citet{1996AJ....111.1705D}	    \\	% UD
       &                  &    5.5$\pm$0.7$^\ast$   & \citet{1998AJ....116..981D}	    \\	% UD
200905 & 5.816$\pm$0.010  &    7.5$\pm$0.6$^\ast$   & \citet{1996AJ....111.1705D}	    \\	% UD
       &                  &    7.5$\pm$0.6$^\ast$   & \citet{1998AJ....116..981D}	    \\	% UD
       &                  &   5.56$\pm$0.04         & \citet{1999AJ....118.3032N}	    \\	
       &                  &   5.61$\pm$0.12         & \citet{2001AJ....122.2707N}	    \\	% NPOI      
       &                  &   5.80$\pm$0.13         & \citet{2001AJ....122.2707N}	    \\	% Mark III
       &                  &  5.787$\pm$0.058        & \citet{2003AJ....126.2502M}	    \\	
208816 & 7.251$\pm$0.012  &       N/A               &                           	    \\	
210745 & 5.302$\pm$0.023  &    5.6$\pm$0.8$^\ast$   & \citet{1998AJ....116..981D}	    \\	% UD
       &                  &   5.32$\pm$0.07         & \citet{2001AJ....122.2707N}	    \\	% NPOI      
       &                  &   5.30$\pm$0.07         & \citet{2001AJ....122.2707N}	    \\	% Mark III
       &                  &  5.234$\pm$0.052        & \citet{2003AJ....126.2502M}	    \\	
213306 & 1.526$\pm$0.018  &   1.63$\pm$0.19         & \citet{1997AandA...317..789M} 	\\	
       &                  &   1.52$\pm$0.02         & \citet{1999AJ....118.3032N}    	\\	
       &                  &  1.520$\pm$0.014        & \citet{2001AJ....121..476A}	    \\	
       &                  &  1.448$\pm$0.007        & \citet{2009MNRAS.394.1925V}	    \\	
214868 & 2.555$\pm$0.047  &   2.63$\pm$0.05         & \citet{1999AJ....118.3032N}   	\\	
       &                  &  2.731$\pm$0.024        & \citet{2010ApJ...710.1365B}	    \\	      
216131 & 2.508$\pm$0.125  &   2.54$\pm$0.06$^\ast$  & \citet{1999AJ....117..521V}	    \\	% UD
       &                  &   2.50$\pm$0.08         & \citet{1999AJ....118.3032N}	    \\	
       &                  &   2.53$\pm$0.09         & \citet{2001AJ....122.2707N}	    \\	% NPOI      
       &                  &   2.49$\pm$0.04         & \citet{2001AJ....122.2707N}	    \\	% Mark III
       &                  &  2.496$\pm$0.040        & \citet{2003AJ....126.2502M}	    \\	
       &                  &  2.369$\pm$0.010        & \citet{2013ApJ...775...45V}	    \\	% UD?
       &                  &  2.457$\pm$0.012        & \citet{2019MNRAS.490.3158C}       \\	
216386 & 8.333$\pm$0.042  &    8.9$\pm$1.0$^\ast$   & \citet{1996AJ....111.1705D}   	\\	% UD
       &                  &    8.9$\pm$0.7$^\ast$   & \citet{1998AJ....116..981D}	    \\	% UD
       &                  &  8.186$\pm$0.105        & \citet{2003AJ....126.2502M}	    \\	
       &                  &   7.77$\pm$1.87$^\ast$  & \citet{2005AandA...434.1201R}	    \\	% UD
       &                  &  7.578$\pm$0.033        & \citet{2019MNRAS.490.3158C}	    \\	
218452 & 1.991$\pm$0.047  &        N/A              &                  	                \\	
224935 & 8.007$\pm$0.059  &    7.2$\pm$0.5$^\ast$   & \citet{1998AJ....116..981D}	    \\	% UD
       &                  &   7.20$\pm$0.70$^\ast$  & \citet{2005AandA...434.1201R}	    \\	% UD
       &                  &  7.245$\pm$0.029        & \citet{2019MNRAS.490.3158C}	        \\	
\enddata
\tablecomments{$^\ast$No LD diameter was provided, therefore we list the UD diameter here. Figure 4 shows a graphical representation of this table. If more than one diameter was available in the literature, we used the most recent one when plotting the results.}
\end{deluxetable*}

%%% ADD AST TO UD DIAMS

\clearpage

%%%%%%%%%%%%%%%%%%%%%%% Figures %%%%%%%%%%%%%%%%%

\begin{figure}[h]
\includegraphics[width=1.0\textwidth]{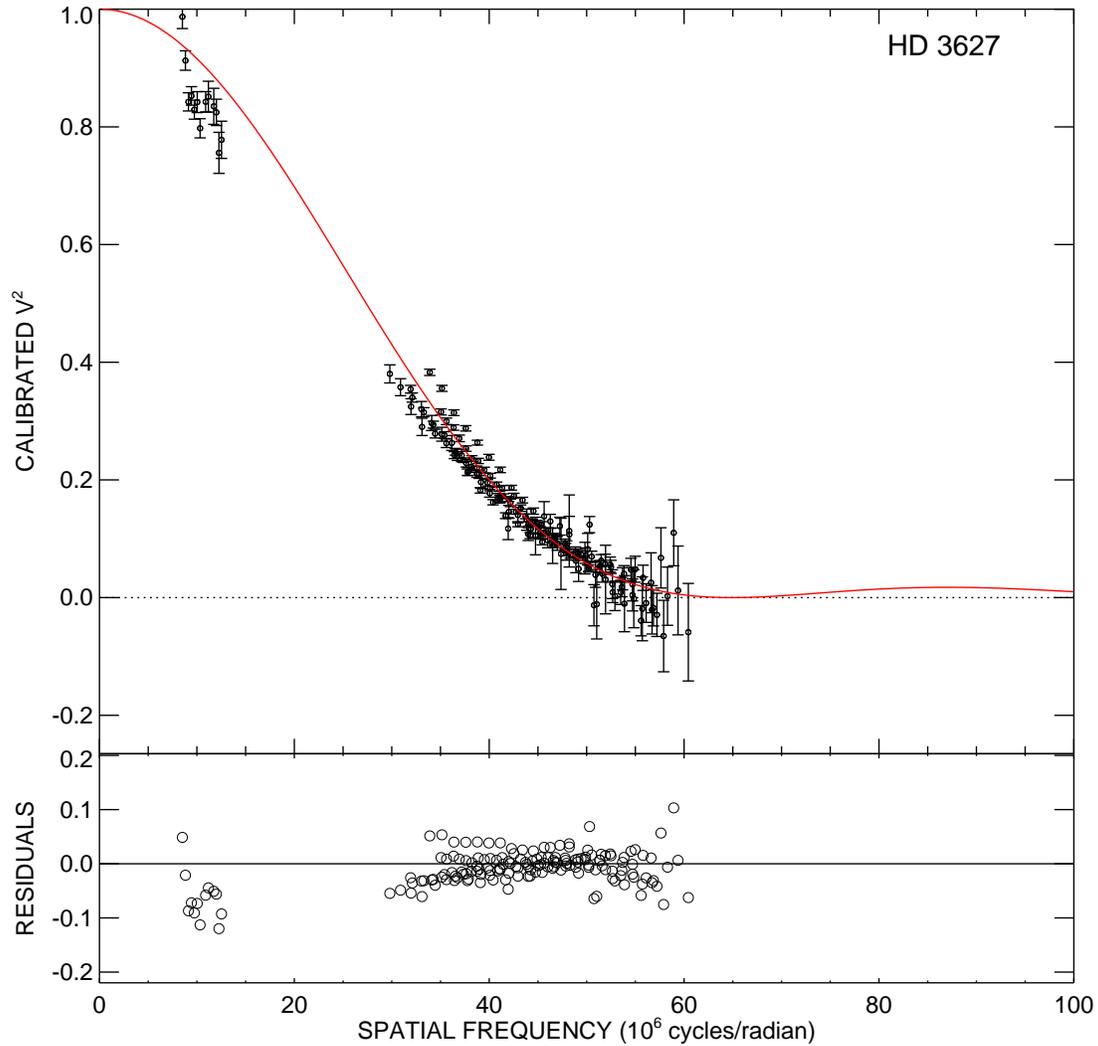}
\caption{\emph{Top panel:} The $\theta_{\rm LD}$ fit for HD 3627 ($\delta$ And). The solid red line represents the visibility curve for the best fit $\theta_{\rm LD}$, the points are the calibrated visibilities, and the vertical lines are the measurement uncertainties. \emph{Bottom panel:} The residuals (O-C) of the diameter fit to the visibilities. The plots for the remaining stars are available on the electronic version of the \emph{Astronomical Journal}.}
  \label{visvsfreq1}
\end{figure}

\clearpage

\begin{figure}[h]
\includegraphics[width=1.0\textwidth]{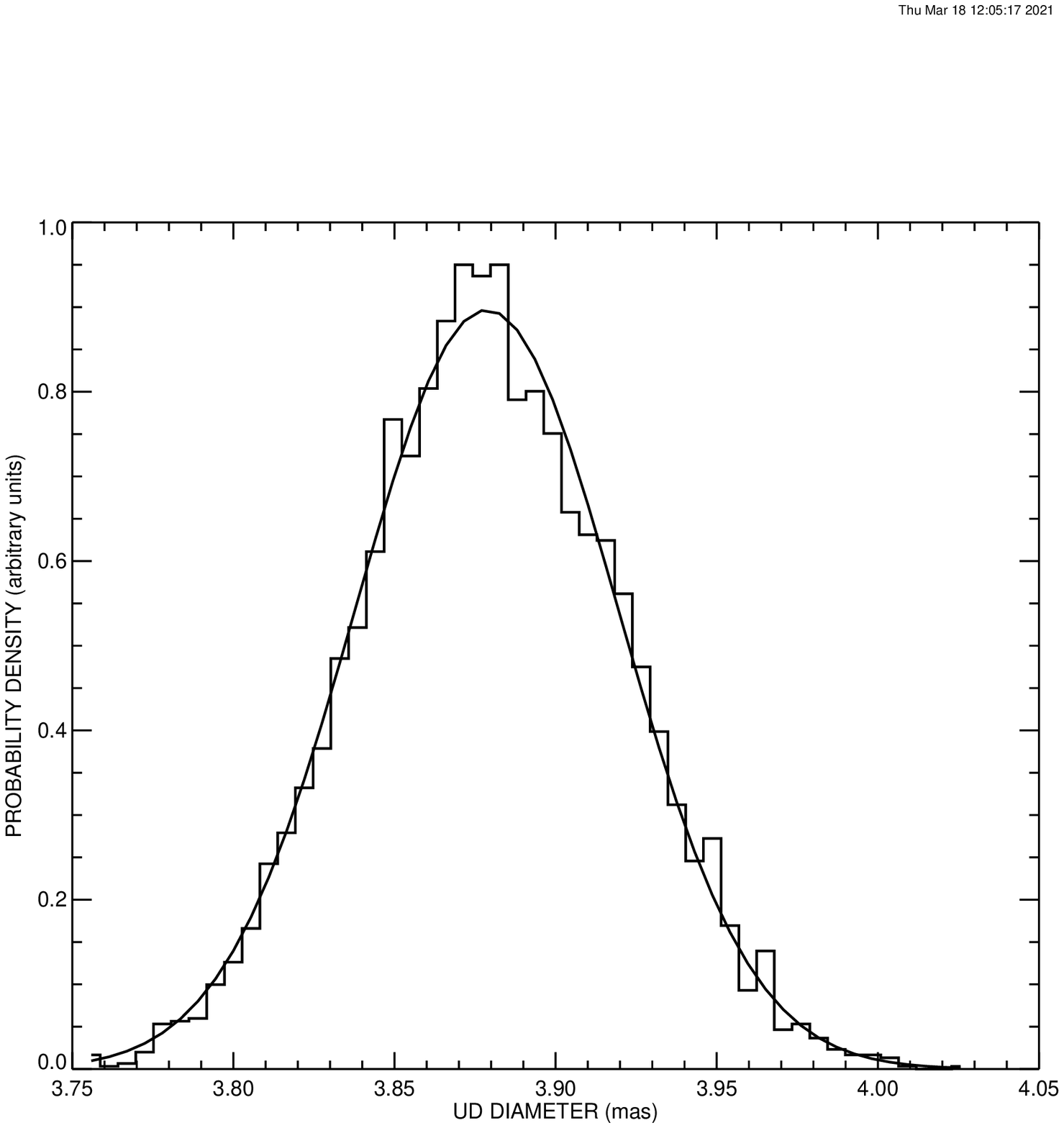}
\caption{An example probability density solution for the diameter fit to HD 3627 ($\delta$ And) visibilities as described in Section 3.1.}
  \label{plot_gauss}
\end{figure}

\clearpage

%\begin{figure}[h]
%\includegraphics[width=1.0\textwidth]{f3.eps}
%\caption{Characterizing the NPOI performance based on the percent error in the limb-darkened diameter measurement ($\sigma_{\rm LD}$) versus $\theta_{\rm LD}$. The horizontal dashed line shows the $\sigma_{\rm LD}$=2$\%$ cut-off that is the minimal standard of astrophysically useful measurements, while the vertical dotted line shows the 3.5-mas cut-off where $\sigma_{\rm LD}$ errors are $\sim 1 \%$ or better. The star with the highest error (HD 120315, $\sigma_{\rm LD} =15\%$) is not included with this plot so that the spread of the other points is more easily visible.}
%  \label{error_compare}
%\end{figure}

%\clearpage

\begin{figure}[h]
\includegraphics[width=1.0\textwidth]{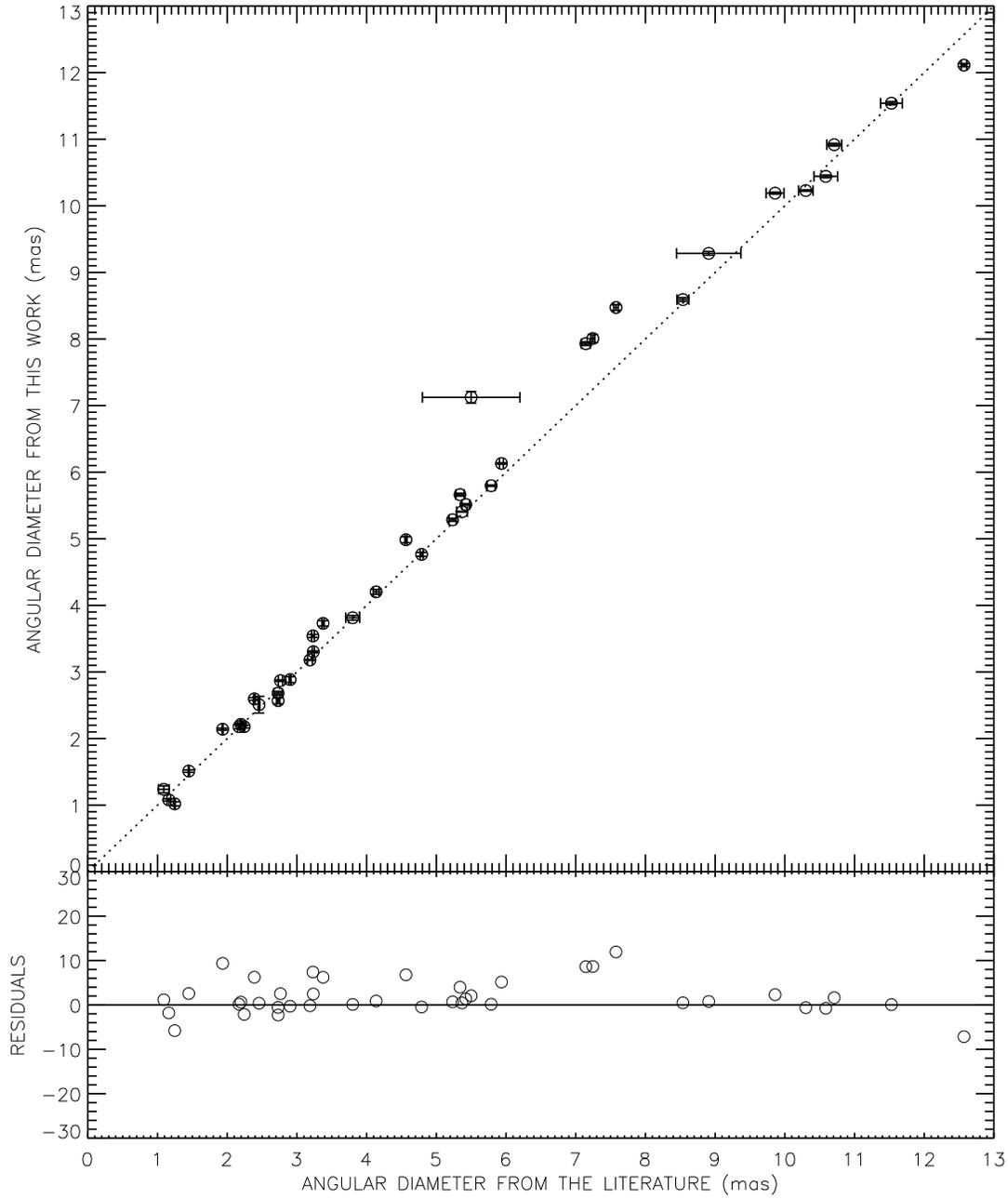}
\caption{\emph{Top panel:} Comparison of the angular diameters measured here versus interferometric diameters from the literature. The error bars for the interferometric diameters are often smaller than the open circle that indicates that measurement. The dotted line is the 1:1 ratio. When more than one measurement was available in the literature, we used the most recent measurement (see Table 7). \emph{Bottom panel:} The residuals were calculated as follows: ($\theta_{\rm here} - \theta_{\rm literature})$ $\times$ (combined error)$^{-1}$.}
  \label{lit_diam_compare}
\end{figure}

\clearpage


\begin{thebibliography}{}

\bibitem[Adelman et al.(2006)]{2006AandA...447..685A} Adelman, S.~J., Caliskan, H., Gulliver, A.~F., et al.\ 2006, \aap, 447, 685

\bibitem[Allende Prieto \& Lambert(1999)]{1999AandA...352..555A} Allende Prieto, C., \& Lambert, D.~L.\ 1999, \aap, 352, 555 

\bibitem[Alonso et al.(1996)]{1996AandAS..117..227A} Alonso, A., Arribas, S., \& Martinez-Roger, C.\ 1996, \aaps, 117, 227 

\bibitem[Anderson \& Francis(2012)]{2012AstL...38..331A} Anderson, E., \& Francis, C.\ 2012, Astronomy Letters, 38, 331 

\bibitem[Armstrong et al.(1998)]{1998ApJ...496..550A} Armstrong, J.~T.,  Mozurkewich, D., Rickard, L.~J, et al.\ 1998, \apj, 496, 550 

\bibitem[Armstrong et al.(2019)]{2019JAI.....850012A} Armstrong, J.~T., Jorgensen, A.~M., Mozurkewich, D., et al.\ 2019, Journal of Astronomical Instrumentation, 8, 1950012-246

\bibitem[Armstrong et al.(2001)]{2001AJ....121..476A} Armstrong, J.~T., Nordgren, T.~E., Germain, M.~E., et al.\ 2001, \aj, 121, 476 

\bibitem[Baines et al.(2014)]{2014ApJ...781...90B} Baines, E.~K., Armstrong, J.~T., Schmitt, H.~R., et al.\ 2014, \apj, 781, 90

\bibitem[Baines et al.(2018)]{2018AJ....155...30B} Baines, E.~K., Armstrong, J.~T., Schmitt, H.~R., et al.\ 2018, \aj, 155, 30

\bibitem[Baines et al.(2010)]{2010ApJ...710.1365B} Baines, E.~K., D{\"o}llinger, M.~P., Cusano, F., et al.\ 2010, \apj, 710, 1365 

\bibitem[Baines et al.(2016)]{2016AJ....152...66B} Baines, E.~K., D{\"o}llinger, M.~P., Guenther, E.~W., et al.\ 2016, \aj, 152, 66 

\bibitem[Baines et al.(2008)]{2008ApJ...680..728B} Baines, E.~K., McAlister, H.~A., ten Brummelaar, T.~A., et al.\ 2008, \apj, 680, 728-733 

\bibitem[Beichman et al.(2018)]{2018AJ....155..158B} Beichman, C.~A., Giles, H.~A.~C., Akeson, R., et al.\ 2018, \aj, 155, 158

\bibitem[Beichman et al.(1988)]{1988iras....1.....B} Beichman, C.~A., Neugebauer, G., Habing, H.~J., et al.\ 1988, Infrared astronomical satellite (IRAS) catalogs and atlases. Volume 1: Explanatory supplement, 1

\bibitem[Benson et al.(2003)]{2003SPIE.4838..358B} Benson, J.~A., Hummel, C.~A., \& Mozurkewich, D.\ 2003, \procspie, 4838, 358 

\bibitem[Bord{\'e} et al.(2002)]{2002AandA...393..183B} Bord{\'e}, P., Coud{\'e} du Foresto, V., Chagnon, G., \& Perrin, G.\ 2002, \aap, 393, 183 

\bibitem[Boyajian et al.(2009)]{2009ApJ...691.1243B} Boyajian, T.~S., McAlister, H.~A., Cantrell, J.~R., et al.\ 2009, \apj, 691, 1243 

\bibitem[Boyajian et al.(2012b)]{2012ApJ...746..101B} Boyajian, T.~S., McAlister, H.~A., van Belle, G., et al.\ 2012b, \apj, 746, 101 

%\bibitem[Boyajian et al.(2015)]{2015MNRAS.447..846B} Boyajian, T., von Braun, K., Feiden, G.~A., et al.\ 2015, \mnras, 447, 846 

\bibitem[Boyajian et al.(2012a)]{2012ApJ...757..112B} Boyajian, T.~S., von Braun, K., van Belle, G., et al.\ 2012a, \apj, 757, 112  

\bibitem[Cardelli et al.(1989)]{1989ApJ...345..245C} Cardelli, J.~A., Clayton, G.~C., \& Mathis, J.~S.\ 1989, \apj, 345, 245

\bibitem[Castelli \& Kurucz(2003)]{2003IAUS..210P.A20C} Castelli, F. \& Kurucz, R.~L.\ 2003, Modelling of Stellar Atmospheres, 210, A20

\bibitem[Cesetti et al.(2013)]{2013AandA...549A.129C} Cesetti, M., Pizzella, A., Ivanov, V.~D., et al.\ 2013, \aap, 549, A129

\bibitem[Charbonnel et al.(2020)]{2020AandA...633A..34C} Charbonnel, C., Lagarde, N., Jasniewicz, G., et al.\ 2020, \aap, 633, A34

\bibitem[Chiavassa et al.(2017)]{2017AandA...600L...2C} Chiavassa, A., Norris, R., Montarg{\`e}s, M., et al.\ 2017, \aap, 600, L2 

\bibitem[Claret \& Bloemen(2011)]{2011AandA...529A..75C} Claret, A., \& Bloemen, S.\ 2011, \aap, 529, A75 

\bibitem[Cohen et al.(2003)]{2003AJ....126.1090C} Cohen, M., Wheaton, W.~A., \& Megeath, S.~T.\ 2003, \aj, 126, 1090 

\bibitem[Colina et al.(1996)]{1996AJ....112..307C} Colina, L., Bohlin, R.~C., \& Castelli, F.\ 1996, \aj, 112, 307 

\bibitem[Cox(2000)]{2000asqu.book.....C} Cox, A.~N.\ 2000, Allen's Astrophysical Quantities (Melville, NY: AIP Press)

\bibitem[Cruzal{\`e}bes et al.(2019)]{2019MNRAS.490.3158C} Cruzal{\`e}bes, P., Petrov, R.~G., Robbe-Dubois, S., et al.\ 2019, \mnras, 490, 3158

\bibitem[Cutri et al.(2003)]{2003yCat.2246....0C} Cutri, R.~M., Skrutskie, M.~F., van Dyk, S., et al.\ 2003, VizieR Online Data Catalog, II/246

\bibitem[di Benedetto \& Conti(1983)]{1983ApJ...268..309D} di Benedetto, G.~P. \& Conti, G.\ 1983, \apj, 268, 309

\bibitem[di Benedetto \& Ferluga(1990)]{1990AandA...236..449D} di Benedetto, G.~P. \& Ferluga, S.\ 1990, \aap, 236, 449

\bibitem[di Benedetto \& Rabbia(1987)]{1987AandA...188..114D} di Benedetto, G.~P., \& Rabbia, Y.\ 1987, \aap, 188, 114 

\bibitem[Dong et al.(2021)]{2021ApJS..255....6D} Dong, J., Huang, C.~X., Dawson, R.~I., et al.\ 2021, \apjs, 255, 6

\bibitem[Ducati(2002)]{2002yCat.2237....0D} Ducati, J.~R.\ 2002, VizieR Online Data Catalog

\bibitem[Duvert et al.(2017)]{2017AandA...597A...8D} Duvert, G., Young, J., \& Hummel, C.~A.\ 2017, \aap, 597, A8 

\bibitem[Dyck et al.(1996)]{1996AJ....111.1705D} Dyck, H.~M., Benson, J.~A., van Belle, G.~T., et al.\ 1996, \aj, 111, 1705

\bibitem[Dyck et al.(1998)]{1998AJ....116..981D} Dyck, H.~M., van Belle, G.~T., \& Thompson, R.~R.\ 1998, \aj, 116, 981 

\bibitem[Eaton \& Williamson(2007)]{2007PASP..119..886E} Eaton, J.~A. \& Williamson, M.~H.\ 2007, \pasp, 119, 886

\bibitem[Fabricius et al.(2002)]{2002AandA...384..180F} Fabricius, C., H{\o}g, E., Makarov, V.~V., et al.\ 2002, \aap, 384, 180.

\bibitem[Faucherre et al.(1983)]{1983AandA...120..263F} Faucherre, M., Bonneau, D., Koechlin, L., et al.\ 1983, \aap, 120, 263

\bibitem[Friedemann(1992)]{1992BICDS..40...31F} Friedemann, C.\ 1992, Bulletin d'Information du Centre de Donnees Stellaires, 40, 31

\bibitem[Gaia Collaboration et al.(2018)]{2018AandA...616A...1G} Gaia Collaboration, Brown, A.~G.~A., Vallenari, A., et al.\ 2018, \aap, 616, A1

\bibitem[Gaia Collaboration(2020)]{2020yCat.1350....0G} Gaia Collaboration\ 2020, VizieR Online Data Catalog, I/350

\bibitem[Gezari et al.(1999)]{1999yCat.2225....0G} Gezari, D.~Y., Pitts, P.~S., \& Schmitz, M.\ 1999, VizieR Online Data Catalog, II/225

\bibitem[Gezari et al.(1993)]{1993cio..book.....G} Gezari, D.~Y., Schmitz, M., Pitts, P.~S., \& Mead, J.~M.\ 1993, Catalog of Infrared Observations, NASA Reference Publication 1294 (3rd ed.; Greenbelt, MD: NASA) 

\bibitem[Gontcharov \& Mosenkov(2018)]{2018yCat.2354....0G} Gontcharov, G.~A. \& Mosenkov, A.~V.\ 2018, VizieR Online Data Catalog, II/354

\bibitem[Gordon et al.(2018)]{2018ApJ...869...37G} Gordon, K.~D., Gies, D.~R., Schaefer, G.~H., et al.\ 2018, \apj, 869, 37

\bibitem[Gordon et al.(2019)]{2019ApJ...873...91G} Gordon, K.~D., Gies, D.~R., Schaefer, G.~H., et al.\ 2019, \apj, 873, 91

\bibitem[Gudennavar et al.(2012)]{2012ApJS..199....8G} Gudennavar, S.~B., Bubbly, S.~G., Preethi, K., et al.\ 2012, \apjs, 199, 8

\bibitem[Hanbury Brown et al.(1974b)]{1974MNRAS.167..121H} Hanbury Brown, R., Davis, J., \& Allen, L.~R.\ 1974b, \mnras, 167, 121 

\bibitem[Hanbury Brown et al.(1967)]{1967MNRAS.137..393H} Hanbury Brown, R., Davis, J., Allen, L.~R., \& Rome, J.~M.\ 1967, \mnras, 137, 393 

\bibitem[Hanbury Brown et al.(1974a)]{1974MNRAS.167..475H} Hanbury Brown, R., Davis, J., Lake, R.~J.~W., \& Thompson, R.~J.\ 1974a, \mnras, 167, 475

\bibitem[Hanke et al.(2018)]{2018AandA...619A.134H} Hanke, M., Hansen, C.~J., Koch, A., et al.\ 2018, \aap, 619, A134

%\bibitem[Heiter et al.(2015)]{2015AandA...582A..49H} Heiter, U., Jofr{\'e}, P., Gustafsson, B., et al.\ 2015, \aap, 582, A49 

\bibitem[Helou \& Walker(1988)]{1988iras....7.....H} Helou, G. \& Walker, D.~W.\ 1988, Infrared astronomical satellite (IRAS) catalogs and atlases. Volume 7, 7, 1

\bibitem[H{\o}g et al.(2000)]{2000AandA...355L..27H} H{\o}g, E., Fabricius, C., Makarov, V.~V., et al.\ 2000, \aap, 355, L27

\bibitem[Hummel et al.(2003)]{2003AJ....125.2630H} Hummel, C.~A., Benson, J.~A., Hutter, D.~J., et al.\ 2003, \aj, 125, 2630

\bibitem[Hummel et al.(1998)]{1998AJ....116.2536H} Hummel, C.~A., Mozurkewich, D., Armstrong, J.~T., et al.\ 1998, \aj, 116, 2536

\bibitem[Hutchings \& Wright(1971)]{1971MNRAS.155..203H} Hutchings, J.~B. \& Wright, K.~O.\ 1971, \mnras, 155, 203

%\bibitem[Hutter \& Elias(2003)]{2003SPIE.4838.1234H} Hutter, D.~J., \& Elias, N.~M., II 2003, \procspie, 4838, 1234

\bibitem[Hutter et al.(2019)]{2019ApJS..243...32H} Hutter, D.~J., Tycner, C., Zavala, R.~T., et al.\ 2019, \apjs, 243, 32

\bibitem[Hutter et al.(2016)]{2016ApJS..227....4H} Hutter, D.~J., Zavala, R.~T., Tycner, C., et al.\ 2016, \apjs, 227, 4 

\bibitem[Jamar et al.(1995)]{1995yCat.3039....0J} Jamar, C., Macau-Hercot, D., Monfils, A., et al.\ 1995, VizieR Online Data Catalog, 3039

\bibitem[Johnson \& Mitchell(1975)]{1975RMxAA...1..299J} Johnson, H.~L. \& Mitchell, R.~I.\ 1975, \rmxaa, 1, 299

\bibitem[Johnson et al.(1966)]{1966CoLPL...4...99J} Johnson, H.~L., Mitchell, R.~I., Iriarte, B., \& Wisniewski, W.~Z.\ 1966, Communications of the Lunar and Planetary Laboratory, 4, 99 

\bibitem[Jones et al.(2016)]{2016ApJ...822L...3J} Jones, J., White, R.~J., Quinn, S., et al.\ 2016, \apjl, 822, L3 

\bibitem[Karata{\textcommabelow s} \& Schuster(2006)]{2006MNRAS.371.1793K} Karata{\textcommabelow s}, Y. \& Schuster, W.~J.\ 2006, \mnras, 371, 1793

\bibitem[Karovicova et al.(2020)]{2020AandA...640A..25K} Karovicova, I., White, T.~R., Nordlander, T., et al.\ 2020, \aap, 640, A25

\bibitem[Kervella et al.(2004)]{2004AandA...413..251K} Kervella, P., Th{\'e}venin, F., Morel, P., et al.\ 2004, \aap, 413, 251

\bibitem[Lafrasse et al.(2010)]{2010SPIE.7734E..4EL} Lafrasse, S., Mella, G., Bonneau, D., et al.\ 2010, \procspie, 7734, 77344E-77344E-11 

\bibitem[Laney et al.(2012)]{2012MNRAS.419.1637L} Laney, C.~D., Joner, M.~D., \& Pietrzy{\'n}ski, G.\ 2012, \mnras, 419, 1637

\bibitem[Ligi et al.(2016)]{2016AandA...586A..94L} Ligi, R., Creevey, O., Mourard, D., et al.\ 2016, \aap, 586, A94 

\bibitem[Marcy et al.(2014)]{2014ApJS..210...20M} Marcy, G.~W., Isaacson, H., Howard, A.~W., et al.\ 2014, \apjs, 210, 20

\bibitem[Mason et al.(2011)]{2011AJ....142..176M} Mason, B.~D., Hartkopf, W.~I., Raghavan, D., et al.\ 2011, \aj, 142, 176

\bibitem[Mason et al.(2001)]{2001AJ....122.3466M} Mason, B.~D., Wycoff, G.~L., Hartkopf, W.~I., et al.\ 2001, \aj, 122, 3466 

\bibitem[McDonald et al.(2017)]{2017MNRAS.471..770M} McDonald, I., Zijlstra, A.~A., \& Watson, R.~A.\ 2017, \mnras, 471, 770

\bibitem[Mermilliod(1991)]{Mermilliod} Mermilliod, J.~C.\ 1991, \emph{Catalogue of Homogeneous Means in the UBV System}, Institut d'Astronomie, Universite de Lausanne

\bibitem[Monet et al.(2003)]{2003AJ....125..984M} Monet, D.~G., Levine, S.~E., Canzian, B., et al.\ 2003, \aj, 125, 984 

\bibitem[Montesinos et al.(2016)]{2016AandA...593A..51M} Montesinos, B., Eiroa, C., Krivov, A.~V., et al.\ 2016, \aap, 593, A51

\bibitem[Mourard et al.(1997)]{1997AandA...317..789M} Mourard, D., Bonneau, D., Koechlin, L., et al.\ 1997, \aap, 317, 789

\bibitem[Mozurkewich et al.(2003)]{2003AJ....126.2502M} Mozurkewich, D., Armstrong, J.~T., Hindsley, R.~B., et al.\ 2003, \aj, 126, 2502 

\bibitem[Mozurkewich et al.(1991)]{1991AJ....101.2207M} Mozurkewich, D., Johnston, K.~J., Simon, R.~S., et al.\ 1991, \aj, 101, 2207 

\bibitem[Neckel et al.(1980)]{1980BICDS..19...61N} Neckel, T., Klare, G., \& Sarcander, M.\ 1980, Bulletin d'Information du Centre de Donnees Stellaires, 19, 61 

\bibitem[Neilson et al.(2017)]{2017ApJ...845...65N} Neilson, H.~R., McNeil, J.~T., Ignace, R., et al.\ 2017, \apj, 845, 65

\bibitem[Ngeow \& Kanbur(2006)]{2006MNRAS.369..723N} Ngeow, C.-C. \& Kanbur, S.~M.\ 2006, \mnras, 369, 723

\bibitem[Nordgren et al.(1999)]{1999AJ....118.3032N} Nordgren, T.~E., Germain, M.~E., Benson, J.~A., et al.\ 1999, \aj, 118, 3032 

\bibitem[Nordgren et al.(2001)]{2001AJ....122.2707N} Nordgren, T.~E., Sudol, J.~J., \& Mozurkewich, D.\ 2001, \aj, 122, 2707 

\bibitem[Oja(1984)]{1984AandAS...57..357O} Oja, T.\ 1984, \aaps, 57, 357

\bibitem[Oja(1993)]{1993AandAS..100..591O} Oja, T.\ 1993, \aaps, 100, 591

\bibitem[Otte \& Dixon(2006)]{2006ApJ...647..312O} Otte, B., \& Dixon, W.~V.~D.\ 2006, \apj, 647, 312 

\bibitem[Pantaleoni Gonz{\'a}lez et al.(2020)]{2020RNAAS...4...12P} Pantaleoni Gonz{\'a}lez, M., Ma{\'\i}z Apell{\'a}niz, J., Barb{\'a}, R.~H., et al.\ 2020, Research Notes of the American Astronomical Society, 4, 12

\bibitem[Perraut et al.(2020)]{2020AandA...642A.101P} Perraut, K., Cunha, M., Romanovskaya, A., et al.\ 2020, \aap, 642, A101

\bibitem[Pickles(1998)]{1998PASP..110..863P} Pickles, A.~J.\ 1998, \pasp, 110, 863

\bibitem[Press et al.(1992)]{1992nrca.book.....P} Press, W.~H., Teukolsky, S.~A., Vetterling, W.~T., \& Flannery, B.~P.\ 1992, Numerical recipes in C. The art of scientific computing (Cambridge: University Press, c1992, 2nd ed.)

\bibitem[Prugniel et al.(2007)]{2007astro.ph..3658P} Prugniel, P., Soubiran, C., Koleva, M., \& Le Borgne, D.\ 2007, arXiv:astro-ph/0703658 

\bibitem[Prugniel et al.(2011)]{2011AandA...531A.165P} Prugniel, P., Vauglin, I., \& Koleva, M.\ 2011, \aap, 531, A165 

\bibitem[Quirrenbach et al.(1993)]{1993ApJ...406..215Q} Quirrenbach, A., Mozurkewich, D., Armstrong, J.~T., et al.\ 1993, \apj, 406, 215

\bibitem[Rains et al.(2020)]{2020MNRAS.493.2377R} Rains, A.~D., Ireland, M.~J., White, T.~R., et al.\ 2020, \mnras, 493, 2377

\bibitem[Richichi \& Percheron(2005)]{2005AandA...434.1201R} Richichi, A. \& Percheron, I.\ 2005, \aap, 434, 1201

\bibitem[Richichi et al.(2009)]{2009MNRAS.399..399R} Richichi, A., Percheron, I., \& Davis, J.\ 2009, \mnras, 399, 399

\bibitem[S{\'a}nchez-Bl{\'a}zquez et al.(2006)]{2006MNRAS.371..703S} S{\'a}nchez-Bl{\'a}zquez, P., Peletier, R.~F., Jim{\'e}nez-Vicente, J., et al.\ 2006, \mnras, 371, 703  

\bibitem[Shao \& Colavita(1992)]{1992ARAandA..30..457S} Shao, M., \& Colavita, M.~M.\ 1992, \araa, 30, 457 

\bibitem[Shao et al.(1988)]{1988ApJ...327..905S} Shao, M., Colavita, M.~M., Hines, B.~E., et al.\ 1988, \apj, 327, 905 

\bibitem[Scowcroft et al.(2016)]{2016MNRAS.459.1170S} Scowcroft, V., Seibert, M., Freedman, W.~L., et al.\ 2016, \mnras, 459, 1170

\bibitem[Smith et al.(2004)]{2004ApJS..154..673S} Smith, B.~J., Price, S.~D., \& Baker, R.~I.\ 2004, \apjs, 154, 673

\bibitem[Soubiran et al.(2016)]{2016AandA...591A.118S} Soubiran, C., Le Campion, J.-F., Brouillet, N., \& Chemin, L.\ 2016, \aap, 591, A118  

\bibitem[Sudol et al.(2002)]{2002AJ....124.3370S} Sudol, J.~J., Benson, J.~A., Dyck, H.~M., et al.\ 2002, \aj, 124, 3370

\bibitem[Swihart et al.(2017)]{2017AJ....153...16S} Swihart, S.~J., Garcia, E.~V., Stassun, K.~G., et al.\ 2017, \aj, 153, 16

\bibitem[Tannirkulam et al.(2008)]{2008SPIE.7013E..0UT} Tannirkulam, A., Monnier, J.~D., Millan-Gabet, R., et al.\ 2008, \procspie, 7013, 70130U

\bibitem[Tycner et al.(2010)]{2010SPIE.7734E.103T} Tycner, C., Hutter, D.~J., \& Zavala, R.~T.\ 2010, \procspie, 7734, 103T

\bibitem[Valdes et al.(2004)]{2004ApJS..152..251V} Valdes, F., Gupta, R., Rose, J.~A., Singh, H.~P., \& Bell, D.~J.\ 2004, \apjs, 152, 251 

\bibitem[van Belle et al.(1999)]{1999AJ....117..521V} van Belle, G.~T., Lane, B.~F., Thompson, R.~R., et al.\ 1999, \aj, 117, 521 

\bibitem[van Belle et al.(2013)]{2013ApJ...775...45V} van Belle, G.~T., Paladini, C., Aringer, B., Hron, J., \& Ciardi, D.\ 2013, \apj, 775, 45 

\bibitem[van Belle et al.(2008)]{2008ApJS..176..276V} van Belle, G.~T., van Belle, G., Creech-Eakman, M.~J., et al.\ 2008, \apjs, 176, 276-292 

\bibitem[van Belle et al.(2009)]{2009MNRAS.394.1925V} van Belle, G.~T., Creech-Eakman, M.~J., \& Hart, A.\ 2009, \mnras, 394, 1925 

\bibitem[van Belle \& von Braun(2009)]{2009ApJ...694.1085V} van Belle, G.~T., \& von Braun, K.\ 2009, \apj, 694, 1085 

\bibitem[van Leeuwen(2007)]{2007AandA...474..653V} van Leeuwen, F.\ 2007, \aap, 474, 653

\bibitem[Wall \& Jenkins(2003)]{2003psa..book.....W} Wall, J.~V., \& Jenkins, C.~R.\ 2003, Practical Statistics for Astronomers, Cambridge Observing Handbooks for Research Astronomers, vol. 3. (Cambridge, UK: Cambridge University Press)

\bibitem[White et al.(2018)]{2018MNRAS.477.4403W} White, T.~R., Huber, D., Mann, A.~W., et al.\ 2018, \mnras, 477, 4403

\bibitem[Wittkowski et al.(2001)]{2001AandA...377..981W} Wittkowski, M., Hummel, C.~A., Johnston, K.~J., et al.\ 2001, \aap, 377, 981 

\bibitem[Wittkowski et al.(2017)]{2017AandA...597A...9W} Wittkowski, M., Arroyo-Torres, B., Marcaide, J.~M., et al.\ 2017, \aap, 597, A9 

\bibitem[Wright(1970)]{1970VA.....12..147W} Wright, K.~O.\ 1970, Vistas in Astronomy, 12, 147

\bibitem[Wu et al.(2011)]{2011AandA...525A..71W} Wu, Y., Singh, H.~P., Prugniel, P., Gupta, R., \& Koleva, M.\ 2011, \aap, 525, A71 

\bibitem[Yee et al.(2017)]{2017ApJ...836...77Y} Yee, S.~W., Petigura, E.~A., \& von Braun, K.\ 2017, \apj, 836, 77

\bibitem[Zacharias et al.(2012)]{2012yCat.1322....0Z} Zacharias, N., Finch, C.~T., Girard, T.~M., et al.\ 2012, VizieR Online Data Catalog, I/322A

\bibitem[Zorec et al.(2009)]{2009AandA...501..297Z} Zorec, J., Cidale, L., Arias, M.~L., et al.\ 2009, \aap, 501, 297 

\end{thebibliography}
\end{document}